\documentclass[preprint,aps,12pt,showpacs,nofootinbib,tightenlines]{revtex4}
\usepackage{amsmath}
\usepackage{amssymb}
\usepackage{epsfig}
\usepackage{graphicx}
\usepackage{subfigure}
\usepackage{color}
\begin{document}
\def\pslash{\rlap{\hspace{0.02cm}/}{p}}
\def\eslash{\rlap{\hspace{0.02cm}/}{e}}
\title {One-loop effects on top pair production in the littlest Higgs model with T-parity at the LHC}
\author{ Bingfang Yang$^1$$^,$$^2$}
\author{ Ning Liu$^1$\footnote{corresponding author: wlln@mail.ustc.edu.cn}}
\affiliation{ $^1$College of Physics $\&$ Electronic Engineering,
Henan Normal University, Xinxiang 453007, China\\ $^2$ Basic
Teaching Department, Jiaozuo University, Jiaozuo 454000, China
   \vspace*{1.5cm}  }

\begin{abstract}

In this work, we systematically investigate the one-loop corrections
to $t\bar{t}$ production in the littlest Higgs model with T-parity
(LHT) at the LHC for $\sqrt{s}=8,14$ TeV. We focus on the effects of
LHT particles on $t\bar{t}$ cross section, polarization asymmetries,
spin correlation and charge asymmetry at the LHC. We also study the
top quark forward-backward asymmetry at Tevatron and its
correlations with the LHC observables. We found that: (1) the
contributions of the LHT particles to $t\bar{t}$ production can only
reach about $1\%$ at the 14 TeV LHC. Meanwhile, the anomalous top
quark forward-backward asymmetry at Tevatron is also hardly to be
explained in the LHT model. (2) the parity violating asymmetries in
$t\bar{t}$ production, such as left-right asymmetry $|A_{LR}|$ and
the polarization $|P_t|$ can respectively reach $1.1\%$ and $0.5\%$,
which may have the potential to provide a signal of LHT at the LHC.

\end{abstract}
\pacs{14.65.Ha,12.15.Lk,12.60.-i,13.85.Lg} \maketitle
\section{ Introduction}
Due to a heavy mass, top quark has been widely considered as a
window of unveiling the new physics at TeV-scale. In particular, the
recent anomalous top quark forward-backward asymmetry observed at
the Tevatron \cite{AFB1} may be a strong hint of new physics beyond
the Standard Model(SM) \cite{AFB2}, although many measurements from
the Tevatron are consistent with the SM predictions. Besides,
because of the small statistics, the study of top-quark properties
is limited at the Tevatron. As a new generation of top quark
factory, the LHC will copiously produce the top events via top pair
productions and single top productions, which provides a good
opportunity to scrutinize the top quark properties and to search the
new physics signals\cite{LHC}.

Notwithstanding the SM have been confirmed by the various
experiments from the LEP to the LHC, it still has some drawbacks,
such as the hierarchy problem. In order to solve this problem, the
little Higgs model was proposed \cite{little Higgs}, where the Higgs
boson is treated as a pseudo-Goldstone boson. The littlest Higgs
(LH) model\cite{LH} is an economical approach to implement the idea
of the little Higgs, however, this model suffers strong constraints
from electro-weak precision tests \cite{constraintLH}and will
reintroduce the fine-tuning problem in the Higgs
potential\cite{Higgs potential}. A feasible way to overcome these
difficulties is to impose a discrete symmetry called
T-parity\cite{T-parity} in the littlest Higgs model, it prevents the
tree-level contribution from the heavy gauge bosons to the
electro-weak observables and also forbids the interactions that
induce the triplet scalars to develop the VEV in the LH model. This
resulting model is referred to as the littlest Higgs model with
T-parity (LHT). One aspect of the LHT phenomenology in top-quark
sector is that the top quark can have interactions with the LHT
particles, such as heavy gauge bosons, mirror fermions, and heavy
quarks $T^{\pm}$. These new interactions can contribute to the
$t\bar{t}$ production at the loop level.

Since the new particles beyond the SM have not been discovered at
the LHC, the scale of the new physics may be higher than the
expected. In this situation, the indirect searches through the loop
effect become important. Furthermore, compared with other light
quarks the produced top quarks can decay before the hadronization,
and the spin information of top quarks will be inherited and
manifested by its decay daughters. Therefore, spin polarization and
spin correlation of top quark can be used to probe the mechanisms of
top quarks productions and decays\cite{top spin SM}, and unveil the
new physics\cite{top spin NP} related to the top quark. In this
paper, we calculate the complete one-loop corrections to the process
$pp\to t\bar{t}$ in the LHT at the Tevatron and at the LHC with
$\sqrt{s}=8$ TeV and 14 TeV. We also study the correlation behaviors
among the top quark forward-backward asymmetry, top charge
asymmetry, polarization asymmetry and the spin correlation.

This paper is organized as follows. In Sec.II we give a brief review
of the LHT model related to our work. In Sec.III we calculate the
(un)polarized $t\bar{t}$ production and the correlation of the
observables in the LHT model. Finally, we give our conclusions in
Sec.IV.

\section{ A brief review of the LHT model}
The LHT is a non-linear $\sigma$ model based on the coset space
$SU(5)/SO(5)$, with the global group $SU(5)$ being spontaneously
broken into $SO(5)$ by a $5\times5$ symmetric tensor at the scale
$f\sim\mathcal{O}(TeV)$, the gauged subgroup $[SU(2)\times
U(1)]^{2}$ of $SU(5)$ is broken into the SM gauge group
$SU(2)_{L}\times U(1)_{Y}$. From the $SU(5)/SO(5)$ breaking, there
arise 14 Goldstone bosons which transform under the electroweak
gauge group as follows:
\begin{equation}
\mathbf{1_0} \oplus \mathbf{3_0} \oplus \mathbf{2_{1/2}} \oplus
\mathbf{3_{\pm1}}.
\end{equation}
where the subscripts indicate the hypercharges. We can denote the
fields in these four representations as $\eta,\omega,H$ and $\phi$,
respectively. After EWSB, $H$ can be decomposed as
$H=(-i\pi^+\sqrt{2},(v+h+i\pi^0)/2)^T$. Explicitly, they are
described by the ``pion" matrix $\Pi$, given by
\begin {equation}
\Pi=
\begin{pmatrix}
-\frac{\omega^0}{2}-\frac{\eta}{\sqrt{20}}&-\frac{\omega^+}{\sqrt{2}}
&-i\frac{\pi^+}{\sqrt{2}}&-i\phi^{++}&-i\frac{\phi^+}{\sqrt{2}}\\
-\frac{\omega^-}{\sqrt{2}}&\frac{\omega^0}{2}-\frac{\eta}{\sqrt{20}}
&\frac{v+h+i\pi^0}{2}&-i\frac{\phi^+}{\sqrt{2}}&\frac{-i\phi^0+\phi^P}{\sqrt{2}}\\
i\frac{\pi^-}{\sqrt{2}}&\frac{v+h-i\pi^0}{2}&\sqrt{4/5}\eta&-i\frac{\pi^+}{\sqrt{2}}&
\frac{v+h+i\pi^0}{2}\\
i\phi^{--}&i\frac{\phi^-}{\sqrt{2}}&i\frac{\pi^-}{\sqrt{2}}&
-\frac{\omega^0}{2}-\frac{\eta}{\sqrt{20}}&-\frac{\omega^-}{\sqrt{2}}\\
i\frac{\phi^-}{\sqrt{2}}&\frac{i\phi^0+\phi^P}{\sqrt{2}}&\frac{v+h-i\pi^0}{2}&-\frac{\omega^+}{\sqrt{2}}&
\frac{\omega^0}{2}-\frac{\eta}{\sqrt{20}}
\end{pmatrix}
\end{equation}
Where $\omega^{\pm},\omega^{0},\eta$ are eaten respectively by 4 new
heavy gauge bosons $W_{H}^{\pm},Z_{H},A_{H}$ whose masses up to
$\mathcal O(\upsilon^{2}/f^{2})$ are given by
\begin {equation}
M_{W_{H}}=M_{Z_{H}}=gf(1-\frac{\upsilon^{2}}{8f^{2}}),M_{A_{H}}=\frac{g'f}{\sqrt{5}}
(1-\frac{5\upsilon^{2}}{8f^{2}})
\end {equation}
with $g$ and $g'$ being the SM $SU(2)$ and $U(1)$ gauge couplings,
respectively. In the 't Hooft-Feynman gauge, the would-be
Goldstone-Boson mass is the same as its corresponding gauge boson.

When T-parity is implemented in the fermion sector of the model we
require the existence of mirror partners for each of the original
fermions. For each SM quark, a copy of mirror quark with T-odd
quantum number is added. We denote them by $u_{H}^{i},d_{H}^{i}$,
where i= 1, 2, 3 are the generation index, whose masses up to
$\mathcal O(\upsilon^{2}/f^{2})$ are given by
\begin{equation}
m_{d_{H}^{i}}=\sqrt{2}\kappa_if, m_{u_{H}^{i}}=
m_{d_{H}^{i}}(1-\frac{\upsilon^2}{8f^2})
\end{equation}
where $\kappa_i$ are the diagonalized Yukawa couplings of the mirror
quarks.

In order to cancel the one-loop quadratic divergent radiative
corrections to Higgs mass parameter induced by top quark, an
additional heavy T-even partner of the top quark $T^{+}$ is
introduced. The implementation of T-parity then requires its own
mirror quark $T^{-}$, which is T-odd under T-parity. Their masses up
to $\mathcal O(\upsilon^{2}/f^{2})$ are given by
\begin{eqnarray}
m_{T^{+}}&=&\frac{f}{v}\frac{m_{t}}{\sqrt{x_{L}(1-x_{L})}}[1+\frac{v^{2}}{f^{2}}(\frac{1}{3}-x_{L}(1-x_{L}))]\\
m_{T^{-}}&=&\frac{f}{v}\frac{m_{t}}{\sqrt{x_{L}}}[1+\frac{v^{2}}{f^{2}}(\frac{1}{3}-\frac{1}{2}x_{L}(1-x_{L}))]
\end{eqnarray}
where $x_{L}$ is the mixing parameter between the SM top-quark $t$
and the new top-quark $T^{+}$.

In the LHT model, the flavor structure is richer than the one of the
SM due to the presence of the mirror fermions and their weak
interactions with the ordinary fermions\cite{LHTflavor}. The mirror
quark sector exists two CKM-like unitary mixing matrices as follows:
\begin{equation}
V_{Hu},V_{Hd}
\end{equation}
Note that $V_{Hu}$ and $V_{Hd}$ are related through the SM CKM
matrix:
\begin{equation}
V_{Hu}^{\dag}V_{Hd}=V_{CKM}.
\end{equation}
These mirror mixing matrices are involved in the flavor changing
interactions between the SM fermions and the mirror fermions which
are mediated by the T-odd gauge bosons ($W_{H}^{\pm},Z_{H},A_{H}$)
or T-odd Goldstone bosons($\omega^{\pm},\omega^{0},\eta$). One
cannot completely turn off the new mixing effects except with a
universally degenerate mass spectrum for the T-odd mirror fermions.
The mixing matrix $V_{Hd}$ can be conveniently parameterized, we
follow Ref.\cite{parameterize} to parameterize $V_{Hd}$ with three
angles $\theta^d_{12},\theta^d_{23},\theta^d_{13}$ and three phases
$\delta^d_{12},\delta^d_{23},\delta^d_{13}$ as follows
\begin{eqnarray}
V_{Hd}=
\begin{pmatrix}
c^d_{12}c^d_{13}&s^d_{12}c^d_{13}e^{-i\delta^d_{12}}&s^d_{13}e^{-i\delta^d_{13}}\\
-s^d_{12}c^d_{23}e^{i\delta^d_{12}}-c^d_{12}s^d_{23}s^d_{13}e^{i(\delta^d_{13}-\delta^d_{23})}&
c^d_{12}c^d_{23}-s^d_{12}s^d_{23}s^d_{13}e^{i(\delta^d_{13}-\delta^d_{12}-\delta^d_{23})}&
s^d_{23}c^d_{13}e^{-i\delta^d_{23}}\\
s^d_{12}s^d_{23}e^{i(\delta^d_{12}+\delta^d_{23})}-c^d_{12}c^d_{23}s^d_{13}e^{i\delta^d_{13}}&
-c^d_{12}s^d_{23}e^{i\delta^d_{23}}-s^d_{12}c^d_{23}s^d_{13}e^{i(\delta^d_{13}-\delta^d_{12})}&
c^d_{23}c^d_{13}
\end{pmatrix}
\end{eqnarray}
\section{Numerical results and discussions}
In our calculation, we neglect the high order $\mathcal
O(\upsilon^{2}/f^{2})$ terms in the masses of new particles and the
higher order couplings between the scalar triplet $\Phi$ and top
quark, the amplitudes are performed at the order $\mathcal
O(\alpha_{s}^{2})$. In the 't Hooft-Feynman gauge, we use the
dimensional regularization scheme to regulate the ultraviolet
divergences in the virtual corrections and adopt the on-shell
renormalization scheme to remove them. The relevant Feynman diagrams
for process $gg \to t\bar{t}$ and $q\bar{q} \to t\bar{t}$ in the LHT
are depicted in Fig.1. The black dot and grey ellipse appearing in
Figs.1 represent the renormalized vertexes
$\hat{\Gamma}^{\mu}_{gt\bar{t}}$, $\hat{\Gamma}^{\mu}_{gq\bar{q}}$
and top quark self-energy at one-loop level respectively, whose
diagrams are displayed in Fig.2 and Fig.3. We list the explicit
expressions of these amplitudes in Appendix. We analytically and
numerically checked that the divergences in the renormalized vertex
and propagator have been canceled. We also find that there are no
divergences in the box diagrams.
\begin{figure}[htbp]
\scalebox{0.42}{\epsfig{file=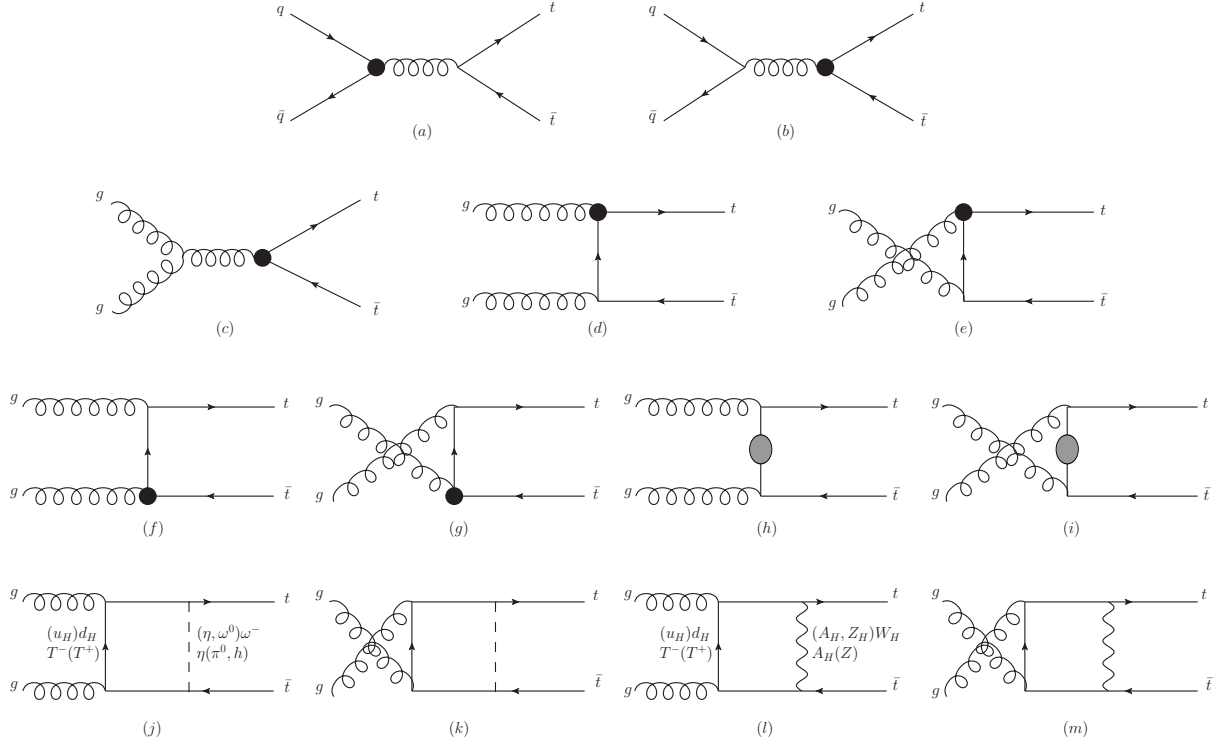}}\caption{Feynman diagrams of
the one-loop correction to the process $pp\rightarrow t\bar{t}$ in
the LHT model. The black dot and grey ellipse represent the
renormalized vertexes $\hat{\Gamma}^{\mu}_{gt\bar{t}}$,
$\hat{\Gamma}^{\mu}_{gq\bar{q}}$ and top quark self-energy
respectively, whose diagrams are displayed in Fig.2 and Fig.3.}
\end{figure}
\begin{figure}[htbp]
\scalebox{0.45}{\epsfig{file=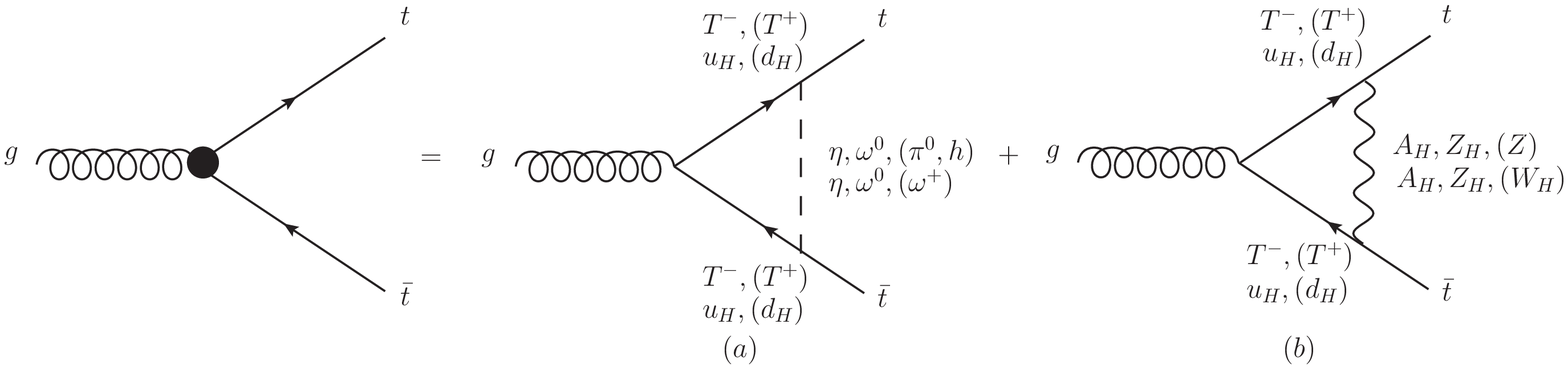}}
\scalebox{0.45}{\epsfig{file=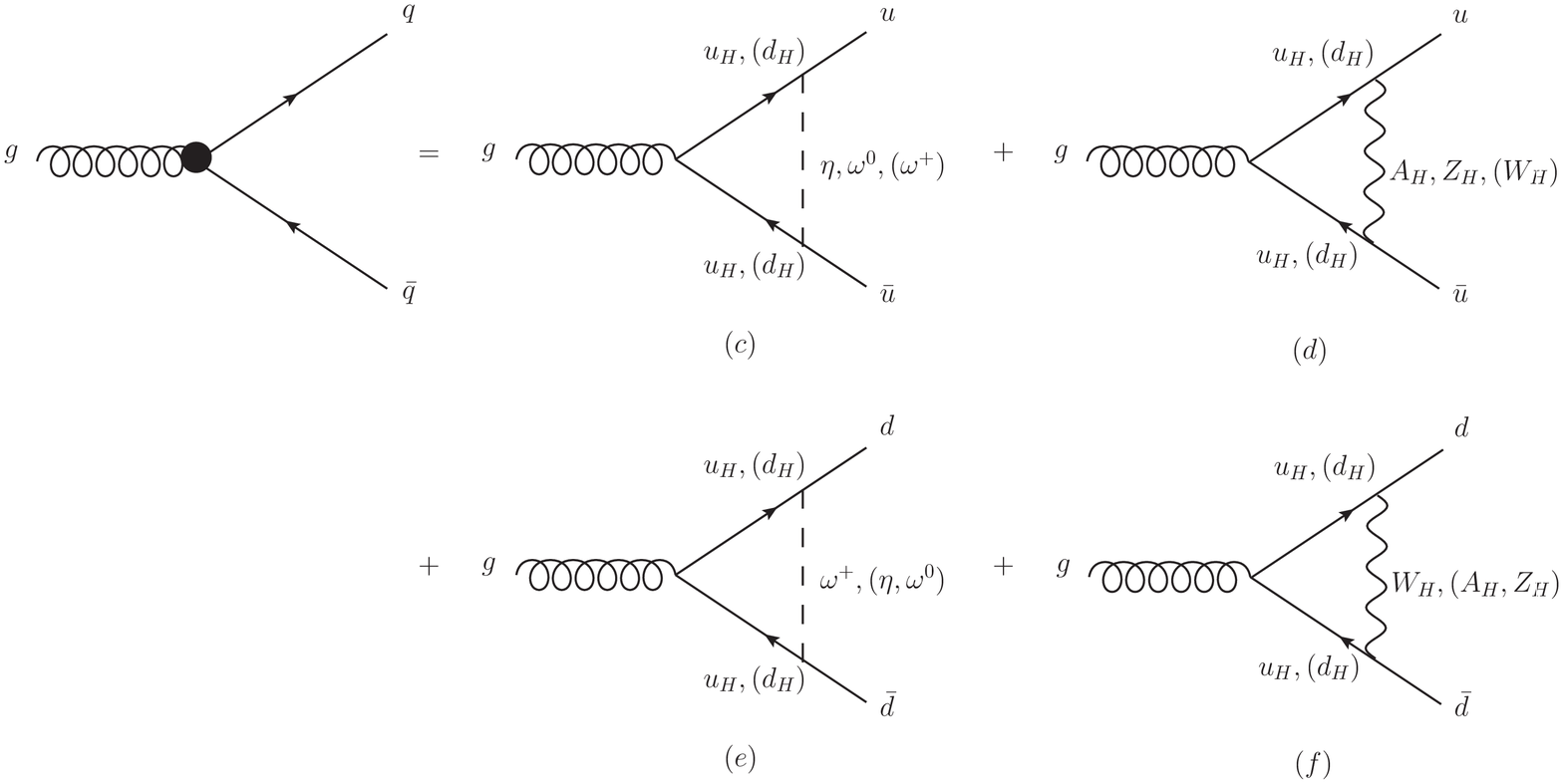}} \caption{The effective $g
t\bar{t},gq \bar{q}$ vertex diagrams at one-loop level in the LHT
model.}
\end{figure}
\begin{figure}[htbp]
\scalebox{0.45}{\epsfig{file=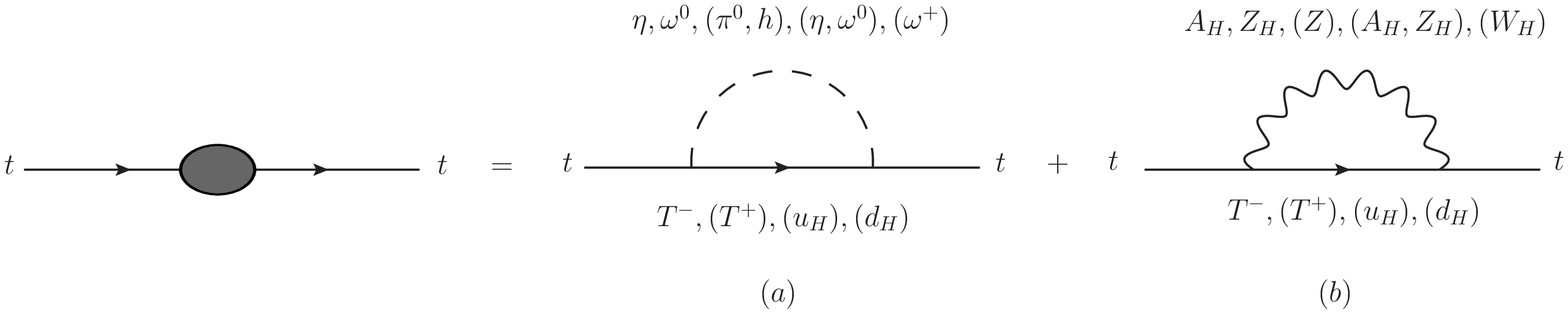}} \caption{The effective fermion
propagator diagrams at one-loop level in the LHT model.}
\end{figure}

The relevant LHT parameters are the scale $f$, the
mixing parameter $x_{L}$, the Yukawa couplings $\kappa_{i}$ and the
parameters in the matrices $V_{H_u},V_{H_d}$. For the mirror fermion
masses, we get $m_{u_{H}^{i}}=m_{d_{H}^{i}}$ at $\mathcal
O(\upsilon/f)$ and assume that the masses of the first two
generations are degeneracy.
\begin{equation}
m_{u_{H}^{1}}=m_{u_{H}^{2}}=m_{d_{H}^{1}}=m_{d_{H}^{2}}=M_{12},m_{u_{H}^{3}}=m_{d_{H}^{3}}=M_{3}
\end{equation}
For the matrices $V_{H_u}$ and $V_{H_d}$, we follow
Ref.\cite{scenario} to choose the following scenario:
$V_{H_u}=1,V_{H_d}=V_{CKM}$. In this scenario, the contribution of
the mirror quarks will come entirely from the third family ones and
the additional heavy quarks $T^{+},T^{-}$. We note that both of the CMS
and ATLAS collaborations have reported their null results of searching for the
fermionic top partner and respectively excluded the masses
regions below 557 GeV \cite{CMS-t}and 656 GeV\cite{ATLAS-t} at 95\%
CL. In our calculations, we scan the parameter
regions: $f=500\sim2000$GeV, $x_{L}=0.1\sim0.9$,
$\sqrt{2}\kappa_{i}=0.6\sim3$ and require our samples to satisfy direct search constraints from the LHC and the
flavor constraints in Refs.\cite{constraint}. Since the new parity violating interactions between top
quark and LHT particles can not only affect the $t\bar{t}$
production rate but also the spin polarization, we will discuss the
LHT corrections to the (un)polarized top pair production by using
the following observables.

\begin{itemize}
\item[(i)] For the unpolarized $t\bar{t}$ production, we calculate the
relative corrections for total $t\bar{t}$ production cross
section($\delta\sigma/\sigma$), charge asymmetry($A_{C}$)\cite{Ac}
at the LHC and the top quark forward-backward
asymmetry($A^{t}_{FB}$)\cite{AFB} at the Tevatron, which are defined as:
\begin{eqnarray}
&&\delta\sigma/\sigma=\frac{\sigma_{tot}-\sigma_{SM}}{\sigma_{SM}},
\\
&&A_{C}=\frac{\sigma(\Delta|\eta_{t}|>0)-\sigma(\Delta|\eta_{t}|
<0)}{\sigma(\Delta|\eta_{t}|>0)+\sigma(\Delta|\eta_{t}|<0)},
\\
&&A_{FB}^{t}=\frac{\sigma(\Delta y_{t}>0)-\sigma(\Delta
y_{t}<0)}{\sigma(\Delta y_{t}>0)+\sigma(\Delta y_{t}<0)}.
\end{eqnarray}
where $\Delta y_t$ ($\Delta\eta_t$) is the (pseudo)rapidity
difference of the top and anti-top quark in the laboratory frame. In
the following, when we calculate $A_C$ and $A_{FB}$, we only
consider the contribution from the interference between the SM and
the LHT model.
\item[(ii)] For the polarized $t\bar{t}$ production, we calculate the
spin correlation($\delta C$)\cite{top spin SM}, the polarization
asymmetry($P_{t}$)\cite{top polarization} and the left-right
asymmetry($A_{LR}$)\cite{top polarization}, which are given by:
\begin{eqnarray}
&&C=\frac{(\sigma_{RR}+\sigma_{LL})-(\sigma_{RL}+\sigma_{LR})}{\sigma_{RR}+\sigma_{LL}+\sigma_{RL}+\sigma_{LR}},\\
&&\delta C=\frac{C_{tot}-C_{SM}}{C_{SM}},\\
&&P_{t}=\frac{(\sigma_{RL}+\sigma_{RR})-(\sigma_{LR}+\sigma_{LL})}{\sigma_{RL}
+\sigma_{RR}+\sigma_{LL}+\sigma_{LR}},\\
&&A_{LR}=\frac{\sigma_{RL}-\sigma_{LR}}{\sigma_{RL}+\sigma_{LR}}.
\end{eqnarray}
Here, the subindices \emph{L(R)} represent
left($\lambda_{t(\bar{t})}=-1/2$) and
right-handed($\lambda_{t(\bar{t})}=+1/2$) top(antitop) quarks,
respectively.
\end{itemize}
The SM parameters input in our numerical calculations are taken
as\cite{parameters}
\begin{eqnarray}
\nonumber G_{F}=1.16637\times10^{-5}\textmd{GeV}^{-2},
s_{W}^{2}=0.231,\alpha_{s}=0.1076,\\
\alpha_{e}=1/128,M_{Z_{L}}=91.2\textmd{GeV},m_{t}=172.9\textmd{GeV}.~~~~~~
\end{eqnarray}
Recently the ATLAS and CMS collaborations at the LHC have
independently discovered a Higgs-like resonance with mass about 125
GeV\cite{Higgsmass}. So we take $m_{h}=125$GeV in our numerical
calculations. We use the parton distribution function
CTEQ10\cite{cteq10} with renormalization scale and factorization
scale $\mu_R = \mu_F = m_t$. In table I, we give the dependence of
our observables on renormalization/factorization scale by taking
$\mu$ as $\mu_0 /2$, $\mu_0$ and $2\mu_0$ respectively. The
benchmark point used in the calculation is: $f=1250$GeV,
$x_{L}=0.5$, $\sqrt{s}=14$TeV (where $A_{FB}^{t}$ for
$\sqrt{s}=1.96$TeV). From the table, we can see that the LHT
corrections will mildly reduce the scale dependence of LO $t\bar{t}$
cross section. For other observables, we find that they have weak
dependence on the unphysical scale because of the cancellation of
scale between numerator and denominator.
\begin{table}[ht]
\begin{center}
\caption[fake]{Dependence of observables in $t\bar{t}$ production on
renormalization/factorization scale with $\mu_0=m_{t}$.}
\bigskip
\begin{tabular}{|c|c|c|c|c|c|c|c|}
\hline  & $\sigma$(pb) & $\delta\sigma/\sigma$(\%) &$A_{C}$(\%)&$A_{FB}^{t}$(\%)&$\delta C$(\%)&$P_{t}$(\%)&$A_{LR}$(\%)\\
\hline  $\mu_0 /2$& 595.14 &$-0.0385$ & $-$0.153
 &$-$0.173 & $-$0.0416 &$-$0.311&$-$0.768
\\
\hline  $\mu_0$& 487.38 &$-0.0385$ & $-$0.154
 &$-$0.173 & $-$0.0415 &$-$0.310&$-$0.770
\\
\hline  $2\mu_0$& 416.18 &$-0.0386$ & $-$0.154
 &$-$0.174 & $-$0.0416 &$-$0.310&$-$0.769
\\
\hline
\end{tabular}
\end{center}
\end{table}
\subsection{Unpolarized top quark pair production}
\begin{figure}[htbp]
\begin{center}
\scalebox{0.7}{\epsfig{file=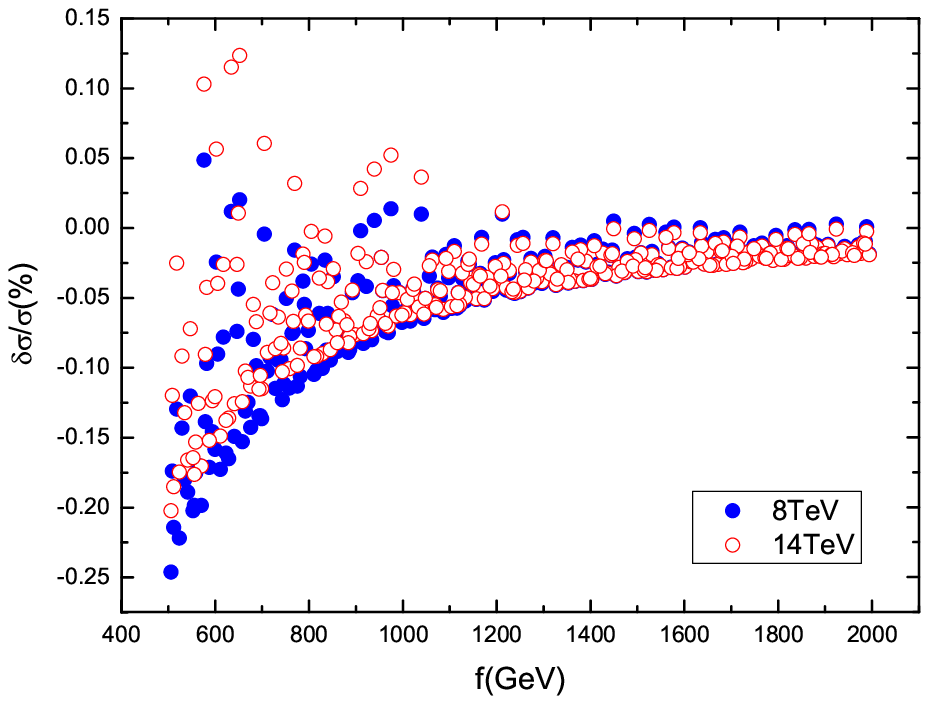}}\vspace{-1cm}
\hspace{-0.5cm} \scalebox{0.7}{\epsfig{file=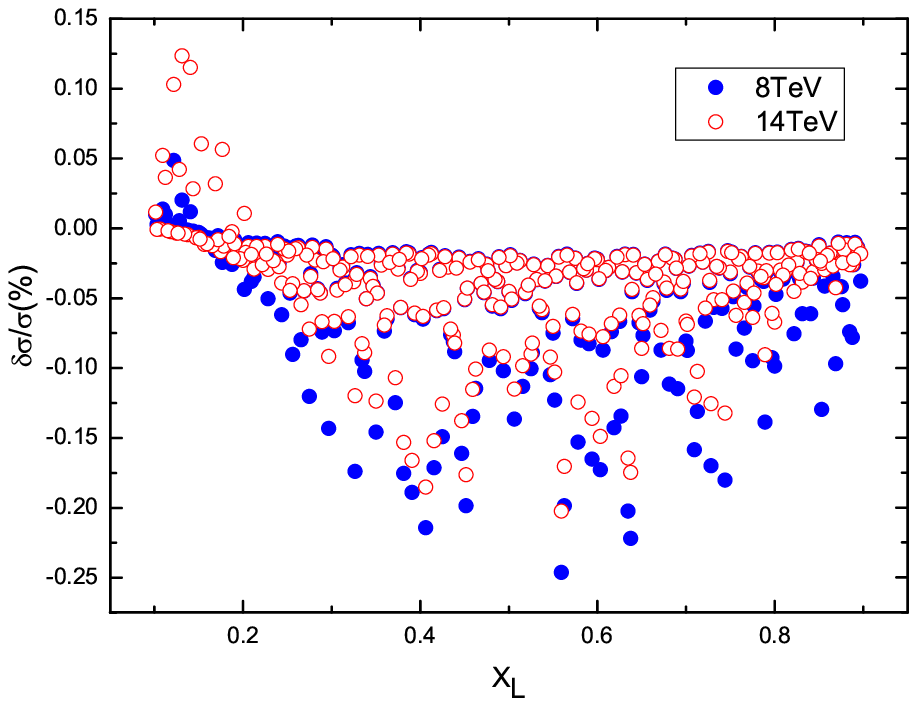}}
 \hspace{-0.5cm} \caption{The relative correction of
the top-quark pair production cross section $\delta \sigma/\sigma$
as the function of $f,x_{L}$ for $\sqrt{s}=8$ TeV and $\sqrt{s}=14$
TeV, respectively.}
\end{center}
\end{figure}

In Fig.4, we show the LHT correction $\delta \sigma/\sigma$
versus $f$ and $x_{L}$ at the LHC when $\sqrt{s}=8,14$ TeV
respectively. On the left panel, we can see that the maximum value of the relative correction to the $t\bar{t}$ cross
section can reach $-0.25$\% for $\sqrt{s}=8$ TeV and $-0.2$\% for
$\sqrt{s}=14$ TeV, respectively. We also notice that the process $gg\rightarrow t\bar{t}$
are more sensitive to the LHT particles than $q\bar{q}\to t\bar{t}$.
When the scale $f$ increases, the relative corrections $\delta \sigma/\sigma$ become small.
This indicates that the effects of the LHT particles on $t\bar{t}$ cross section will decouple at the high cutoff scale $f$.
Since heavy top quark $T^{+}$ and $T^{-}$ masses have a strong dependence on the mixing parameter $x_{L}$, we can see that
when $x_{L}$ tends to 0, the masses of $T^{+}$ and $T^{-}$ will
become heavy and their contribution is very small. When $x_{L}$
tends to 1, the masses of $T^{+}$ will become heavy but the masses
of $T^{-}$ will become light. As a result, the effect of $T^{-}$ will still reside in the $t\bar{t}$ production. On the right
panel of Fig.4, we can see that the maximum value of the relative
correction $\delta \sigma/\sigma$ occurs in the region of $x_{L}\sim 0.56$.

\begin{figure}[htbp]
\begin{center}
\scalebox{0.7}{\epsfig{file=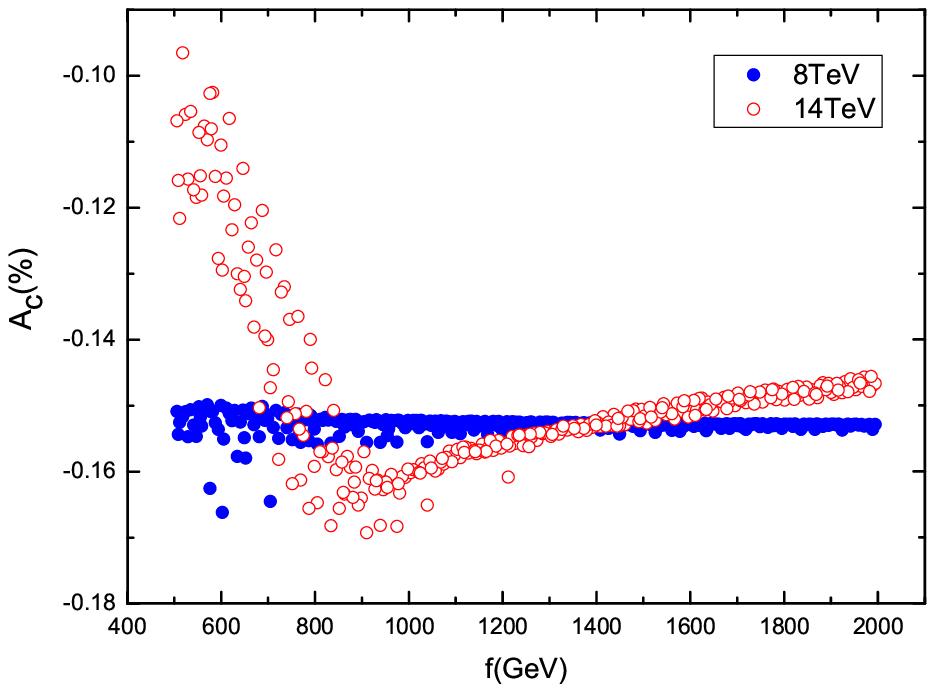}} \vspace{-1cm} \hspace{-0.5cm}
\scalebox{0.7}{\epsfig{file=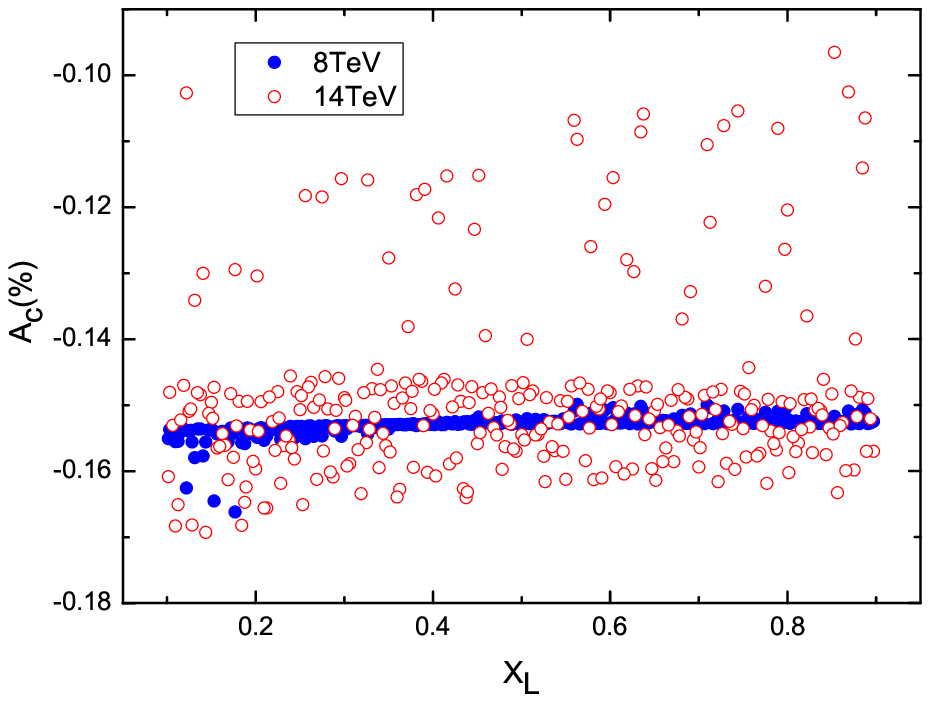}}
 \hspace{-0.5cm} \caption{The top quark charge
asymmetry $A_{C}(t\bar{t})$ as the function of $f,x_{L}$ for
$\sqrt{s}=8$ TeV and $\sqrt{s}=14$ TeV, respectively. The
$A_{C}(t\bar{t})$ plotted therein correspond to the LHT
contributions.}
\end{center}
\end{figure}

In Fig.5 we show the charge asymmetry $A_{C}(t\bar{t})$ versus $f$
and $x_{L}$ at the LHC, where $A_{C}(t\bar{t})$ only includes the
LHT contributions. We can see that the LHT contribution to charge
asymmetry $A_{C}(t\bar{t})$ are negative and small for
$\sqrt{s}=8,14$ TeV. Considering the uncertainty of $t\bar{t}$
measurement, we can infer that it will be very difficult to observe
the LHT effects on $A_C(t\bar{t})$ at the LHC\cite{AcSM,Ac}. We also
notice that since both numerator and denominator of $A_C$ in Eq.(13)
decouple with the cutoff scale $f$, $A_C(t\bar{t})$ has a weak
dependence on the scale $f$ and show a slow decoupling behavior. We
checked that when $f$ was taken very large, the LHT effects on
$A_C(t\bar{t})$ will disappear. The similar behavior can be seen in
$A^{t}_{FB}$ in Fig.6.

\begin{figure}[htbp]
\begin{center}
\scalebox{0.7}{\epsfig{file=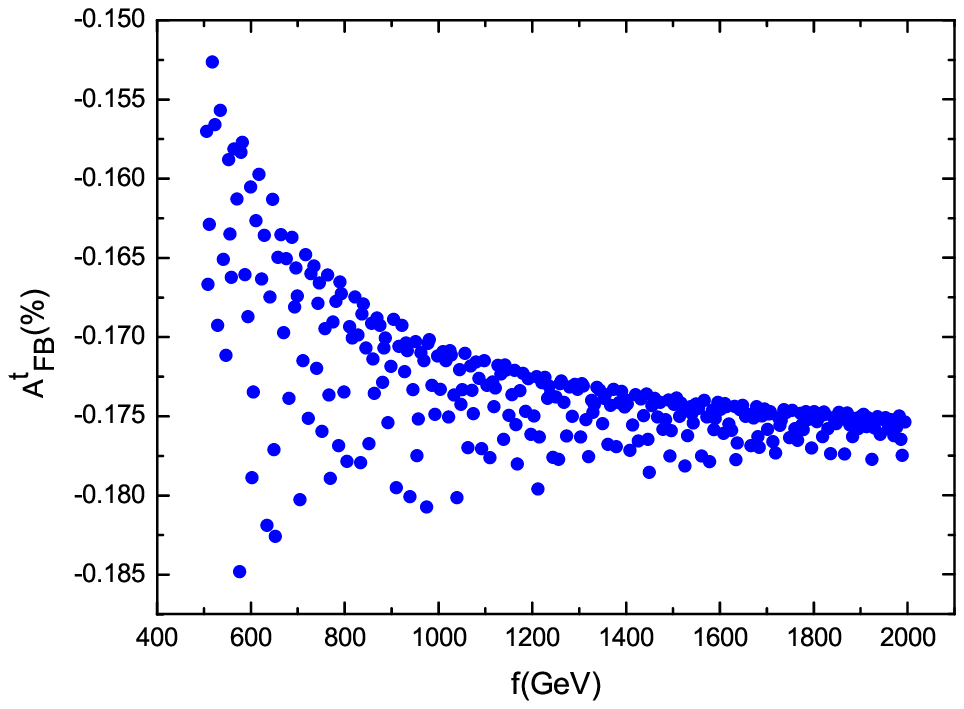}}\hspace{-0.5cm}
\scalebox{0.7}{\epsfig{file=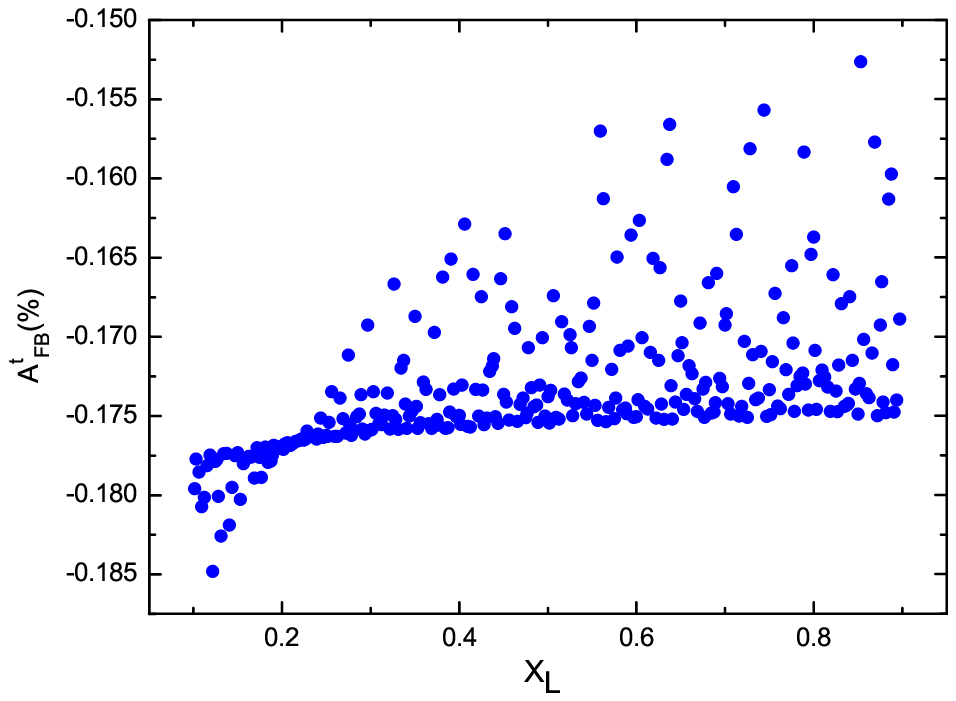}} \vspace{-1cm} \caption{The
forward-backward asymmetry in rapidity $A_{FB}^{t}$ as the function
of $f,x_{L}$ at the Tevatron. The $A_{FB}^{t}$ plotted therein
correspond to the LHT contributions.}
\end{center}
\end{figure}

In Fig.6 we show the forward-backward asymmetry $A_{FB}^{t}$ versus
$f,x_{L}$ at the Tevatron, where $A_{FB}^{t}$ only includes the LHT
contributions. We can see $A_{FB}^{t}$ in the LHT is negative and
small. So LHT model will be not helpful to alleviate the large
discrepancy between the SM prediction and the measurement of
$A^{t}_{FB}$ from Tevatron.

\subsection{Polarized top quark pair production}
\begin{itemize}

\item[(i)]The correction to the spin correlation

\begin{figure}[htbp]
\begin{center}
\scalebox{0.7}{\epsfig{file=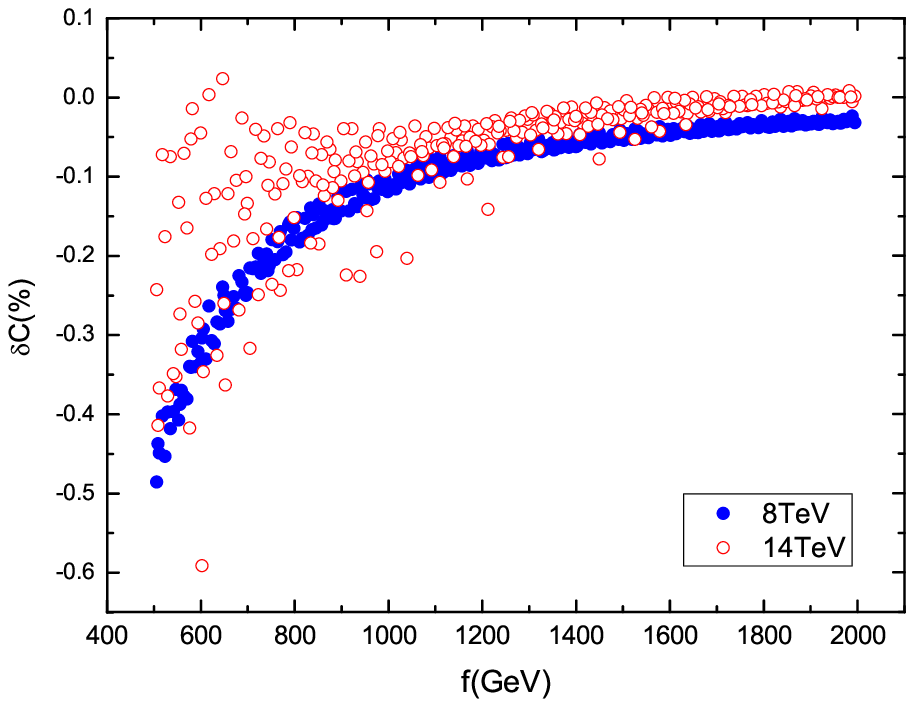}}
\scalebox{0.7}{\epsfig{file=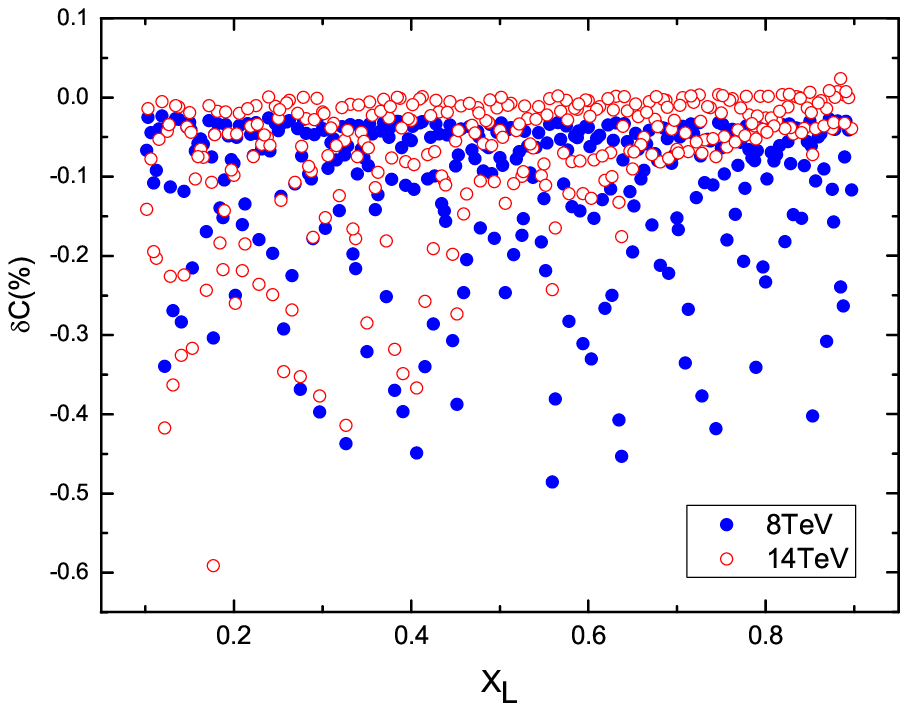}}\vspace{-1cm} \caption{The
correction to the spin correlation $\delta C$ as the function of
$f,x_{L}$ for $\sqrt{s}=8$ TeV and $\sqrt{s}=14$ TeV, respectively.}
\end{center}
\end{figure}
Recently, the CMS collaboration reported their measurement of the
$t\bar{t}$ spin correlation coefficient $C$, that is, $C=0.24\pm
0.02(stat.)\pm 0.08(syst.)$ in the helicity basis\cite{dcLHC2},
which is consistent with the SM predictions. In Fig.7, we show the
relative correction to the spin correlation $\delta C$ versus
$f,x_{L}$ for the LHC with $\sqrt{s}=8,14$ TeV, respectively. We can
see $\delta C$ decouple fast with the increase of scale $f$. The
maximum value of $\delta C$ can reach $-0.5\%$ for $\sqrt{s}=8$ TeV
and $-0.6\%$ for $\sqrt{s}=14$ TeV, which is difficult to be
detected at the LHC\cite{dcLHC1}.

\item[(ii)] Top quark polarization asymmetry

\begin{figure}[htbp]
\begin{center}
\scalebox{0.7}{\epsfig{file=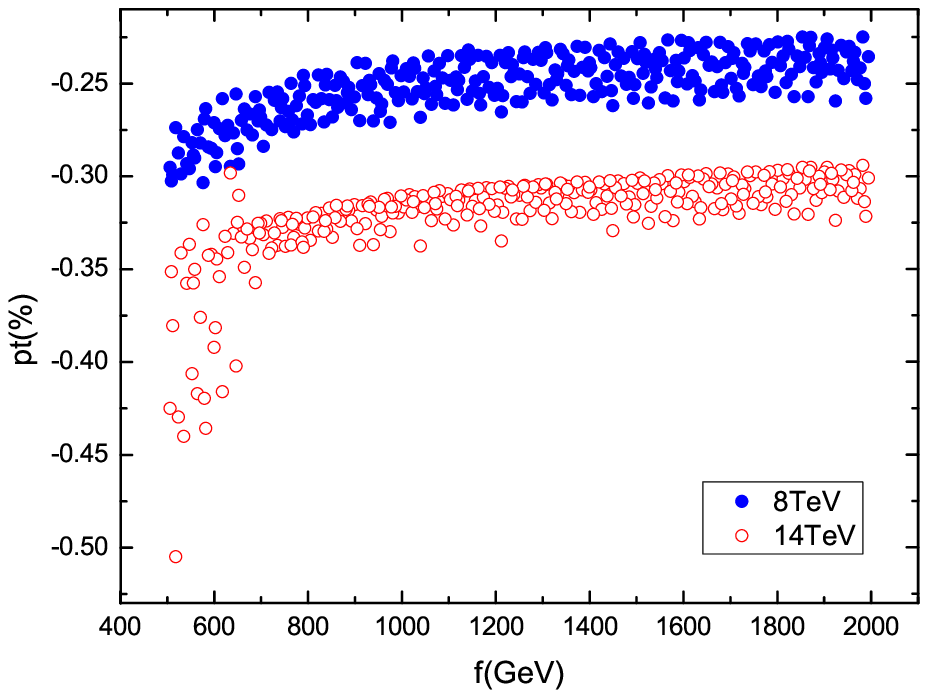}}
\scalebox{0.7}{\epsfig{file=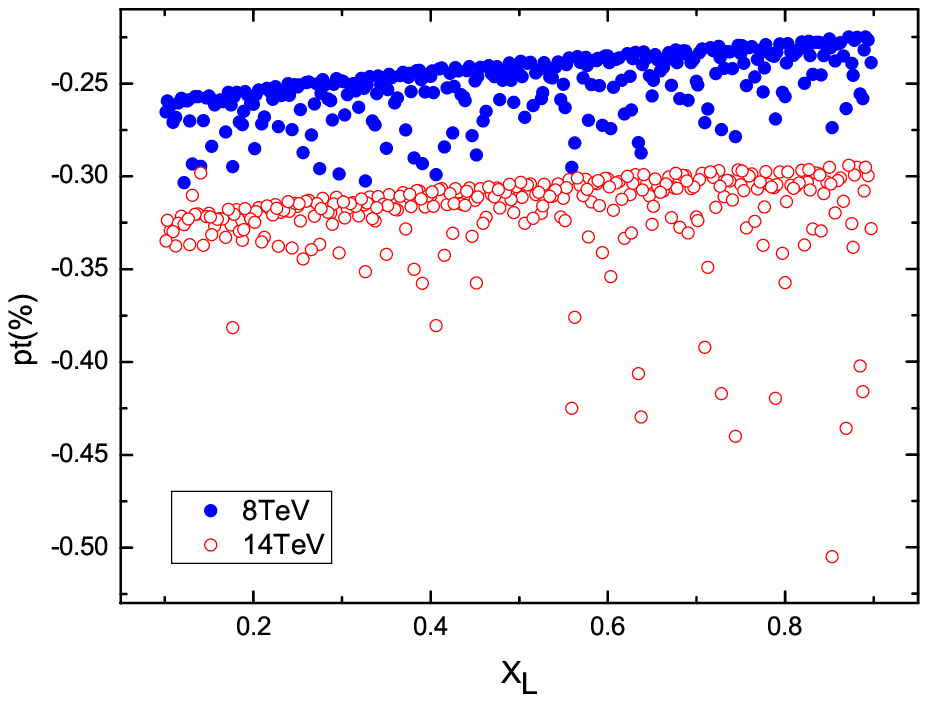}}\vspace{-1cm} \caption{The top
quark polarization asymmetry $P_{t}$ as the function of $f,x_{L}$
for $\sqrt{s}=8$ TeV and $\sqrt{s}=14$ TeV, respectively.}
\end{center}
\end{figure}

In Fig.8 we show the polarization asymmetry $P_{t}$ versus $f,x_{L}$
for the LHC with $\sqrt{s}=8,14$ TeV, respectively. From Fig.8, we
can see the large effects come from the region of small $f$ and
large $x_{L}$. Comparing with the results of $\sqrt{s}=8$ TeV, we
can see that the $P_{t}$ is enhanced greatly for $\sqrt{s}=14$ TeV.
The maximum value of $P_{t}$ can respectively reach about $-0.3\%$
for $\sqrt{s}=8$ TeV and about $-0.5\%$ for $\sqrt{s}=14$ TeV.
Compared to $|P_{t}|=0.5\%$ predicted in the SM at the 14 TeV
LHC\cite{ALR}, the $t\bar{t}$ polarization asymmetries in the LHT
model may be accessible at the LHC.

\item[(iii)] Top quark left-right asymmetry

\begin{figure}[htbp]
\begin{center}
\scalebox{0.7}{\epsfig{file=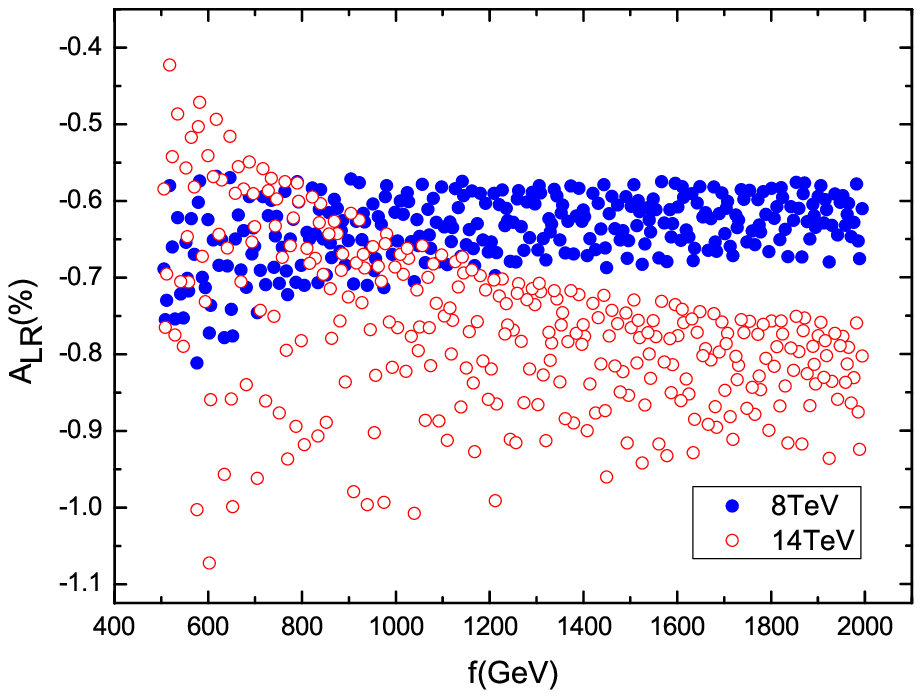}}
\scalebox{0.7}{\epsfig{file=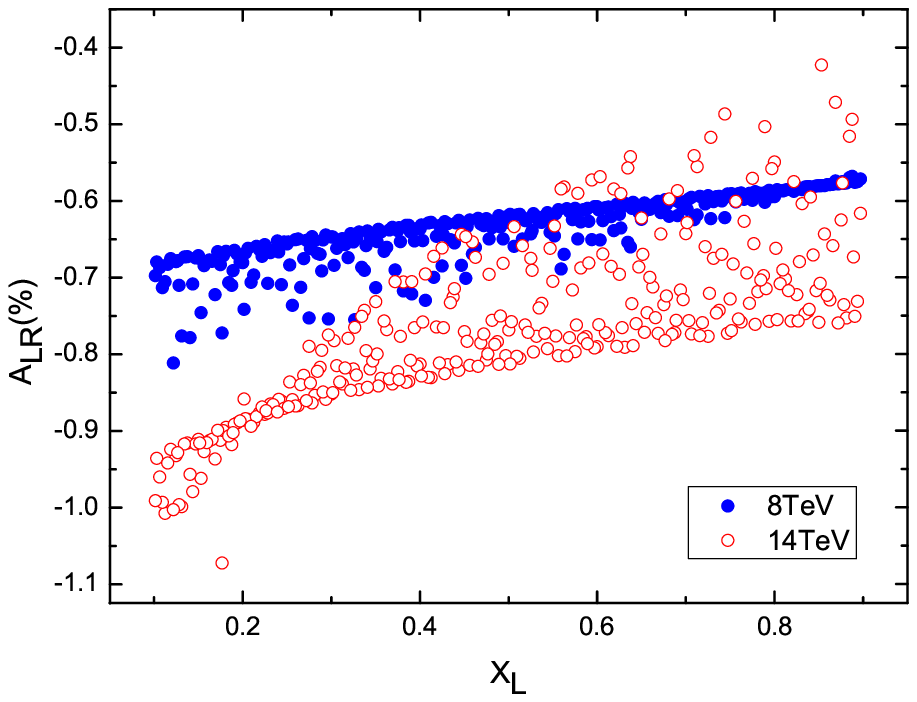}}\vspace{-1cm} \caption{The
top quark left-right asymmetry $A_{LR}$ as the function of $f,x_{L}$
for $\sqrt{s}=8$ TeV and $\sqrt{s}=14$ TeV, respectively. The
$A_{LR}$ plotted therein correspond to the LHT contributions.}
\end{center}
\end{figure}

In Fig.9, we show the left-right asymmetry $A_{LR}$ versus $f,x_{L}$
for the LHC with $\sqrt{s}=8,14$ TeV. We can see that
$A_{LR}$ can maximally reach about $-0.8\%$ and about
$-1.1\%$ for $\sqrt{s}=8,14$ TeV respectively. In order to estimate the statistical
observability of $A_{LR}$, we use the significance $N_{S}$ defined in Refs.\cite{top
polarization} and find that the maximal significance of $A_{LR}$ can be $3\sigma$ for $\sqrt{s}=8$ TeV and $9.3\sigma$ for
$\sqrt{s}=14$ TeV with an assumption of integrated luminosity $\cal{L}$ $=5.0 fb^{-1}$.
 It has also been pointed in Ref.\cite{top polarization} the appropriate cuts on $t\bar{t}$ invariant mass and on the angle between the lepton and top quark can further enhance the significance at the LHC. Therefore, the left-right asymmetries in $t\bar{t}$ production may have the potential to probe LHT at the LHC and may deserve further studies by including the top quark deay and detector response.
\end{itemize}

\subsection{The correlation of the observables in $t\bar{t}$ production}
\begin{figure}[htbp]
\begin{center}
\scalebox{0.7}{\epsfig{file=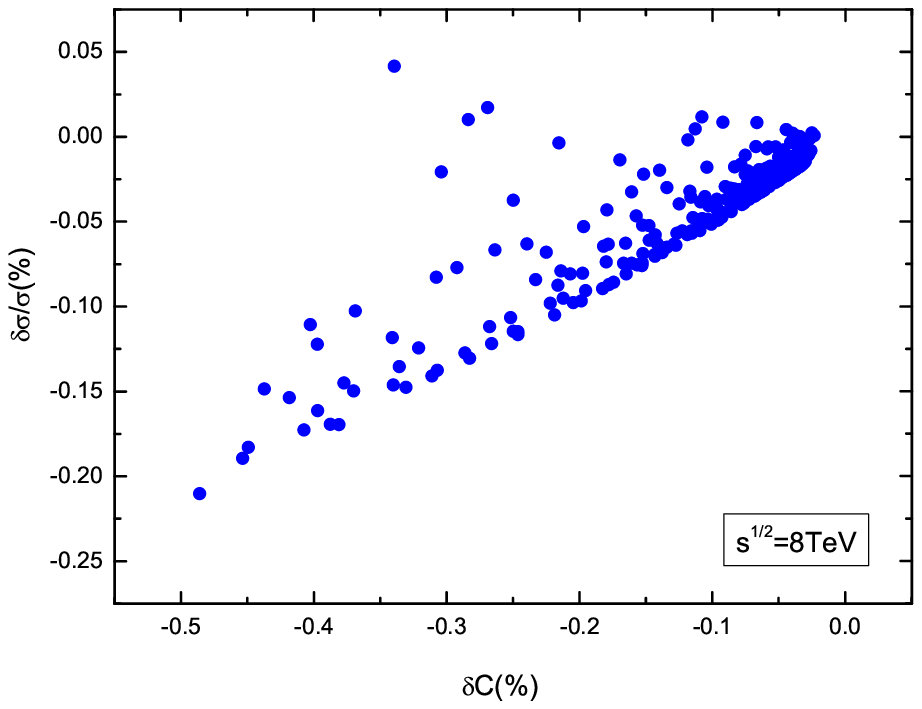}}\vspace{-1cm}
\hspace{-0.5cm} \scalebox{0.7}{\epsfig{file=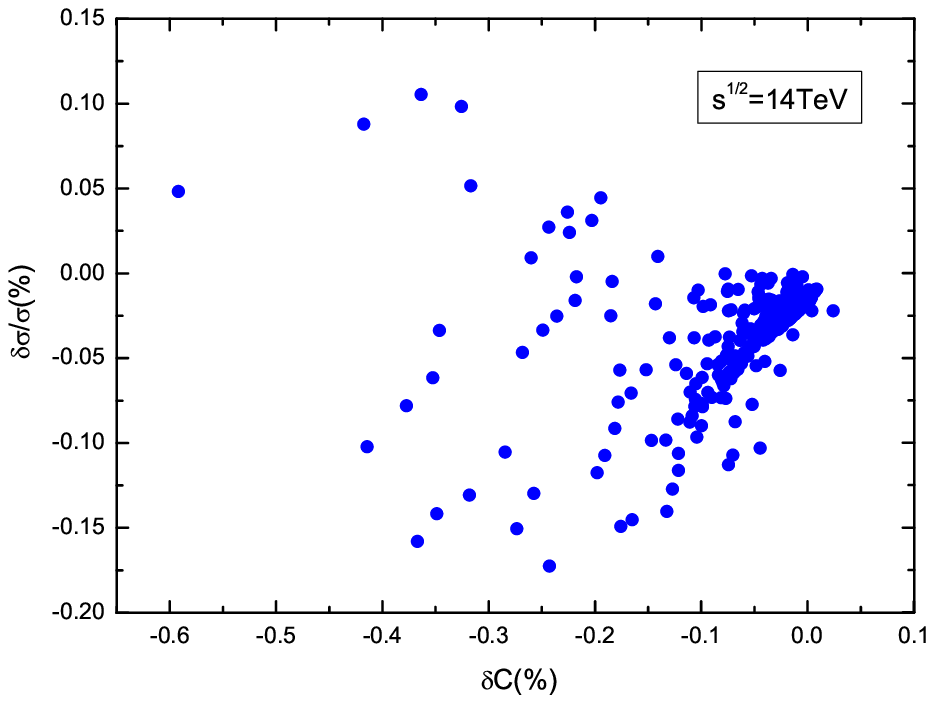}}
\scalebox{0.7}{\epsfig{file=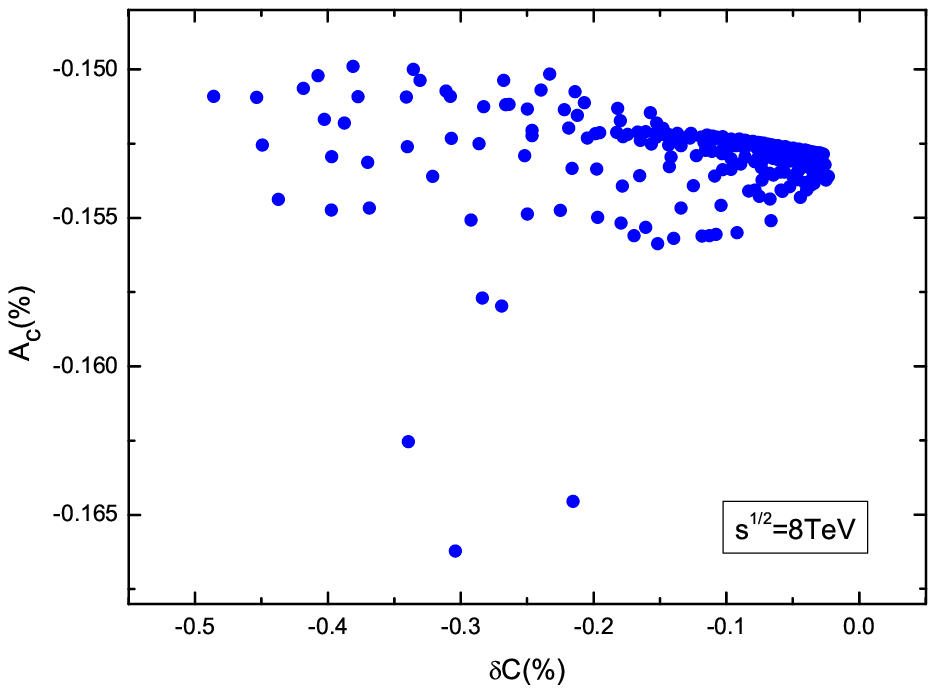}}\vspace{-1cm}
\hspace{-0.5cm} \scalebox{0.7}{\epsfig{file=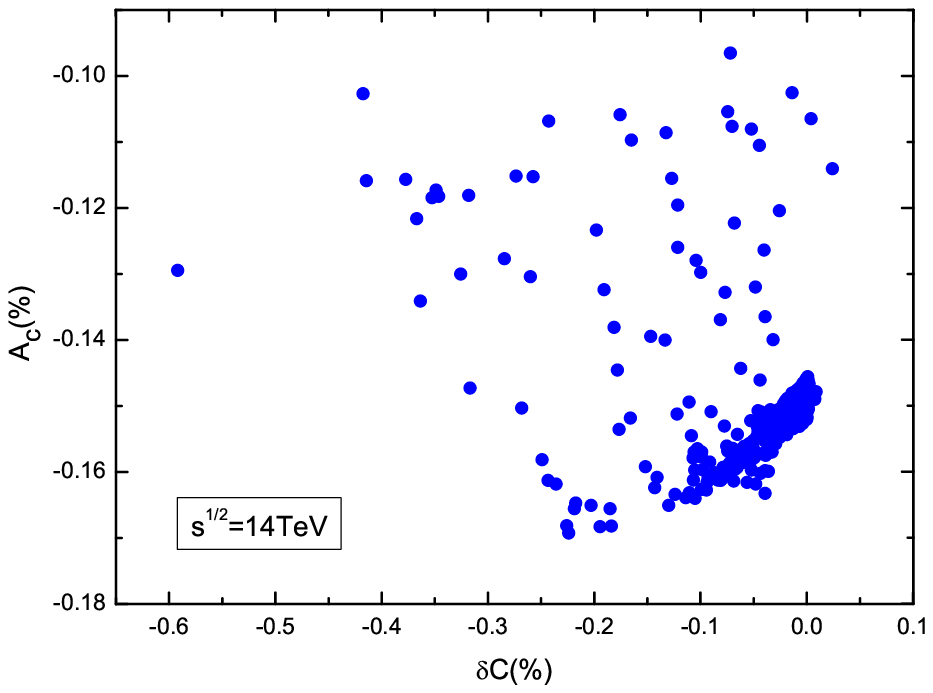}}
\scalebox{0.7}{\epsfig{file=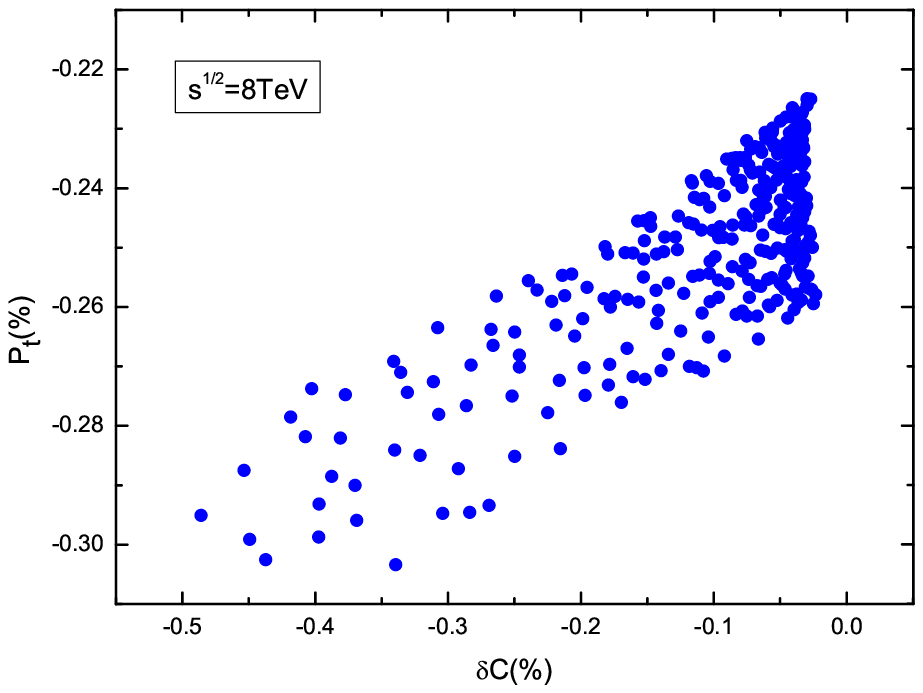}}\vspace{-1cm}
\hspace{-0.5cm} \scalebox{0.7}{\epsfig{file=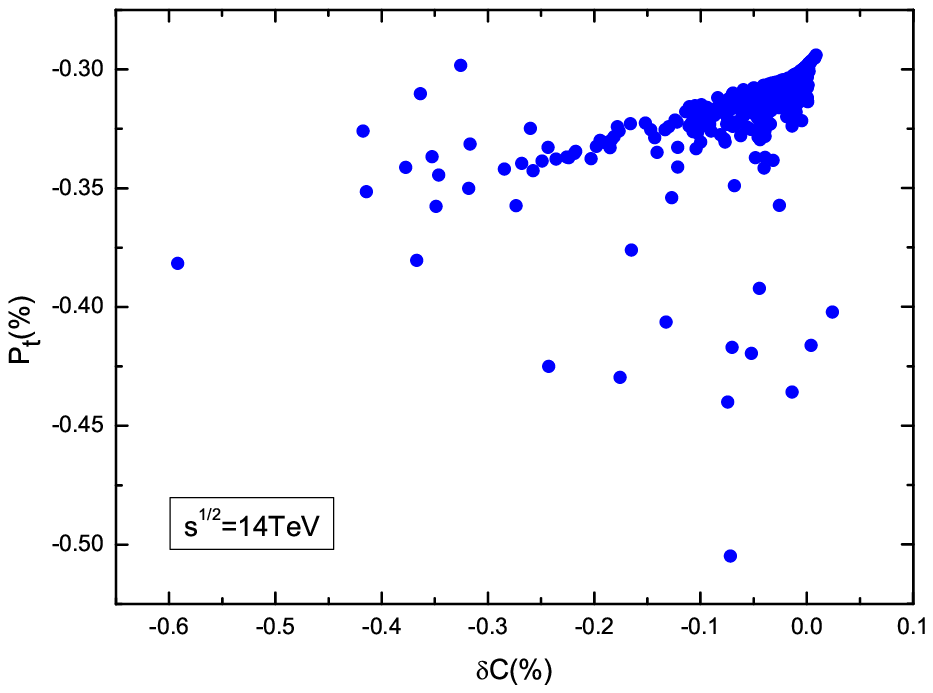}}
\scalebox{0.7}{\epsfig{file=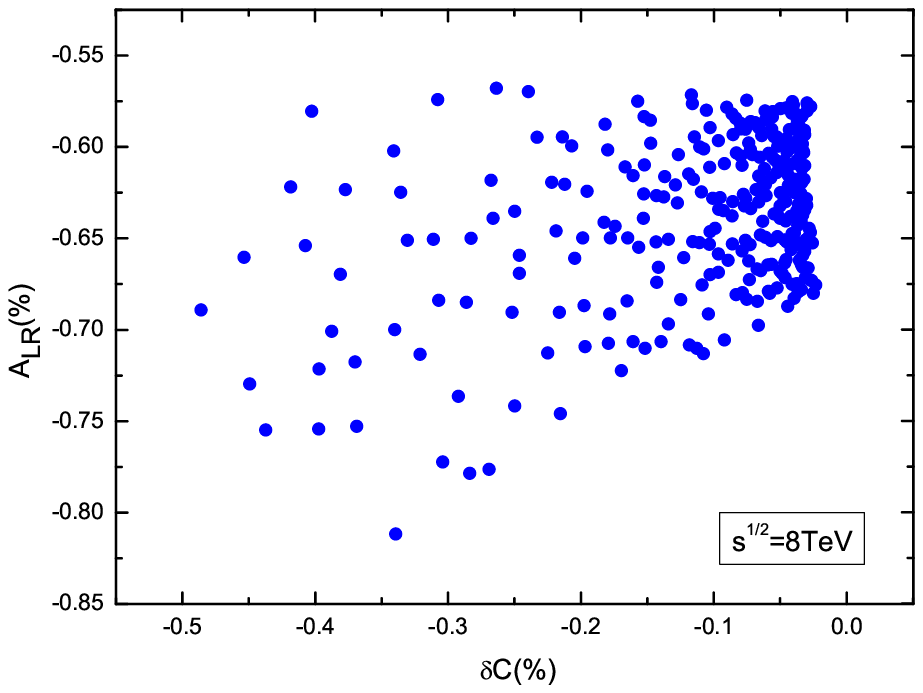}}\vspace{-1cm}
\hspace{-0.3cm}\scalebox{0.7}{\epsfig{file=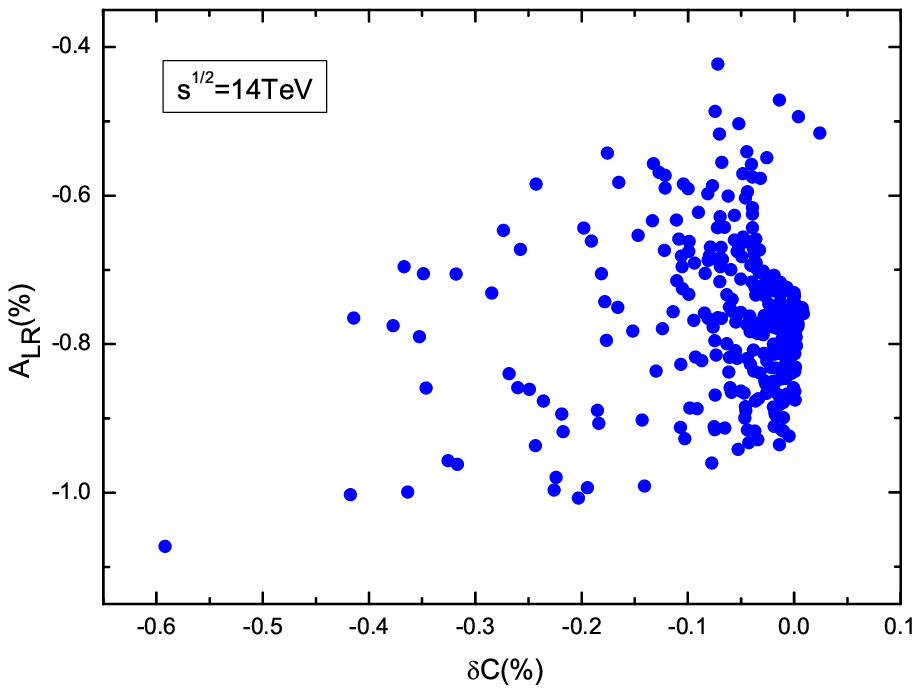}}
\caption{The correlations between $\delta C$ and $\delta
\sigma/\sigma, A_{C}(t\bar{t}), P_{t}, A_{LR}$ for $\sqrt{s}=8$ TeV
and $\sqrt{s}=14$ TeV, respectively.}
\end{center}
\end{figure}
In Fig.10, we present the correlations among $\delta C$, $\delta
\sigma/\sigma, A_{C}(t\bar{t}), P_{t}$ and $A_{LR}$ at the LHC with
$\sqrt{s}=8$ TeV and $\sqrt{s}=14$ TeV. We can see that the
correlation behaviors of these observables at $\sqrt{s}=8$ TeV are
similar to the ones at $\sqrt{s}=14$ TeV. Since the new chiral
interactions can simultaneously affect $\delta C$, $P_{t}$ and
$A_{LR}$, we can see that there is strong correlation among the
three observables. Besides, we notice that correlation among $\delta
C$, $P_{t}$ and $A_{LR}$ in LHT model are different from those in
other new physics models, such as axigluon model, left-right
symmetric models\cite{wulei polarizaiotn paper}. So we can use these
correlation behaviors to distinguish the LHT model from other new
physics models.

\section{Conclusions} \noindent

 In this paper, we systematically studied the one-loop LHT corrections to $t\bar{t}$ production
 at the LHC for $\sqrt{s}=8,14$ TeV. We presented the numerical results for the relative correction
 to $t\bar{t}$ cross section, polarization asymmetries, spin correlation and charge asymmetry at the LHC.
 Besides, we also investigated the top quark forward-backward asymmetry at Tevatron and its correlations
 with the LHC observables. We found that the effects of the LHT particles are significant only when they
 are light and the largest relative correction from these particles to $t\bar{t}$ production can only
 reach about $1\%$. So it will be difficult to observe such small loop-induced LHT effects through the
 measurement of $t\bar{t}$ cross section at the LHC. Meanwhile, the anomalous top quark forward-backward
 asymmetry at Tevatron is also hardly to be explained in the LHT model. However, we noticed that the
 contribution from LHT to left-right asymmetry $|A_{LR}|$ and the polarization $|P_t|$ can respectively
 reach $1.1\%$ and $0.5\%$, compared to $|A_{LR}|=1.2\%$\cite{ALR} and $|P_{t}|=0.5\%$ within the SM at the 14 TeV LHC.
 These parity violating asymmetries in $t\bar{t}$ production may have the potential to probe LHT at the LHC
 and may deserve further studies by including the top quark decay and detector response and optimizing the
 cuts on $t\bar{t}$ invariant mass and on the angle between the lepton and top quark to enhance the significance at the LHC.

\textbf{Acknowledgments}\\
We would like to thank Lei Wu for useful discussions and
suggestions. This work is supported by the National Natural Science
Foundation of China (NNSFC) under grant No.11305049, the Startup
Foundation for Doctors of Henan Normal University under contract
No.11112, the National Natural Science Foundation of China (Grant
No. 11147167), the key project of science and technology research of
the Education Department of Henan province under Grant Nos.
13A140113 and 12A140011, the Program of ISTTCP under Grant No.
114100510021, and the Natural Science Research Project of Education
Department of Henan Province (Grant No. 2011A140007).
\begin{center}
\textbf{Appendix: The explicit expressions of the renormalized
vertex $\hat{\Gamma}^{\mu}_{gf\bar{f}}$ and the renormalized
propagator $-i\hat{\Sigma}^f(p)$} \cite{renormalization}
\end{center}

They can be represented in form of 1-point, 2-point and 3-point
standard functions $A,B_{0},B_{1},C_{ij}$. Here $p_{t}$ and $p'_{t}$
denote the momenta of the top and antitop respectively, and they are
assumed to be outgoing.

(I)Renormalization vertex
\begin{figure}[htbp]
\scalebox{0.4}{\epsfig{file=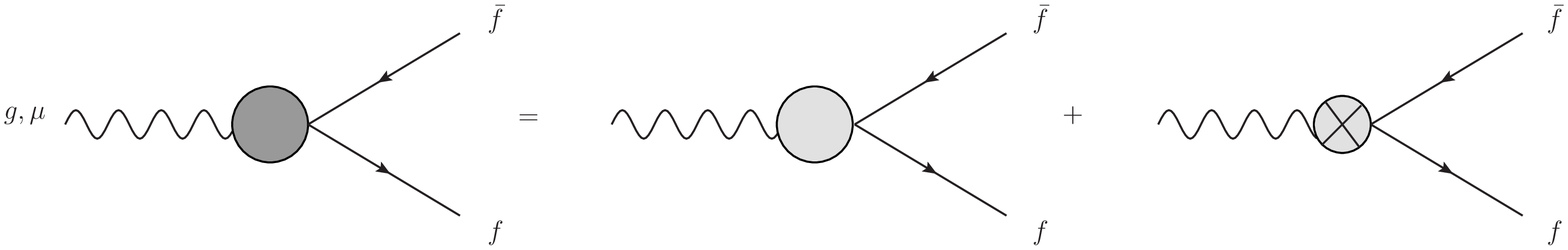}}
\end{figure}
\begin{eqnarray}
\hat{\Gamma}^{\mu}_{gf\bar{f}}&=&\Gamma^{\mu}_{g
f\bar{f}}-ieQ_{f}\gamma^{\mu}(\delta Z_{V}^{f}-\gamma_{5}\delta
Z_{A}^{f}-\frac{S_{W}}{2C_{W}}\delta Z_{ZA})
+ie\gamma^{\mu}(v_{f}-a_{f}\gamma_{5})\frac{1}{2}\delta
Z_{ZA}\nonumber
\end{eqnarray}

where
\begin{eqnarray*}
&&v_{f}\equiv\frac{I_{f}^{3}-2Q_{f}S_{W}^{2}}{2C_{W}S_{W}},\quad
a_{f}\equiv\frac{I_{f}^{3}}{2C_{W}S_{W}}\\
&&\delta Z_{ZA}=2\frac{\Sigma_{T}^{AZ}(0)}{M_{Z_{L}}^{2}}\\
&&\delta
Z_{L}^{f}=Re\Sigma_{L}^{f}(m_{f}^{2})+m_{f}^{2}\frac{\partial}{\partial
P_{f}^{2}}Re[\Sigma_{L}^{f}(P_{f}^{2})+\Sigma_{R}^{f}(P_{f}^{2})+2\Sigma_{S}^{f}(P_{f}^{2})]|_{P_{f}^{2}=m_{f}^{2}}\\
&&\delta
Z_{R}^{f}=Re\Sigma_{R}^{f}(m_{f}^{2})+m_{t}^{2}\frac{\partial}{\partial
P_{f}^{2}}Re[\Sigma_{L}^{f}(P_{f}^{2})+\Sigma_{R}^{f}(P_{f}^{2})+2\Sigma_{S}^{f}(P_{f}^{2})]|_{P_{f}^{2}=m_{f}^{2}}\\
&&\delta Z_{V}^{f}=\frac{1}{2}(\delta Z_{L}^{f}+\delta
Z_{R}^{f}),\delta Z_{A}^{f}=\frac{1}{2}(\delta Z_{L}^{f}-\delta
Z_{R}^{f})
\end{eqnarray*}
\begin{eqnarray*}
\hat{\Gamma}^{LHT,\mu}_{gt\bar{t}}&=&\hat{\Gamma}^{\mu}_{g
t\bar{t}}(\eta)+ \hat{\Gamma}^{\mu}_{g
t\bar{t}}(\omega^{0})+\hat{\Gamma}^{\mu}_{g
t\bar{t}}(\omega^{\pm})+\hat{\Gamma}^{\mu}_{g t\bar{t}}(\pi^{0})
+\hat{\Gamma}^{\mu}_{g t\bar{t}}(h)\\&+&\hat{\Gamma}^{\mu}_{g
t\bar{t}}(A_{H})+\hat{\Gamma}^{\mu}_{g t\bar{t}}(Z_{H})+
\hat{\Gamma}^{\mu}_{g t\bar{t}}(W_{H}^{\pm})+\hat{\Gamma}^{\mu}_{g
t\bar{t}}(Z)
\end{eqnarray*}
\begin{eqnarray}
&&\hat{\Gamma}^{\mu}_{u_{H}\eta}=\frac{g'^{2}g_{s}T^{a}_{\alpha\beta}(V_{Hu})_{i3}^{*}(V_{Hu})_{i3}}{100M_{A_{H}}^{2}}\frac{i}{16\pi^{2}}\nonumber\\
&&\{m_{u_{H}}^{2}\gamma^{\alpha}\gamma^{\mu}\gamma^{\beta}C_{\alpha\beta}P_{L}+m_{u_{H}}^{2}(\pslash'_{t}+\pslash_{t})\gamma^{\mu}\gamma^{\alpha}C_{\alpha}P_{L}
+m_{t}^{2}\gamma^{\alpha}\gamma^{\mu}\gamma^{\beta}C_{\alpha\beta}P_{R}\nonumber\\
&&+m_{t}^{2}(\pslash'_{t}+\pslash_{t})\gamma^{\mu}\gamma^{\alpha}C_{\alpha}P_{R}
-m_{u_{H}}^{2}m_{t}\gamma^{\alpha}\gamma^{\mu}C_{\alpha}P_{R}-m_{u_{H}}^{2}m_{t}(\pslash'_{t}+\pslash_{t})\gamma^{\mu}C_{0}P_{R}\nonumber\\
&&-m_{u_{H}}^{2}m_{t}\gamma^{\alpha}\gamma^{\mu}C_{\alpha}P_{L}-m_{u_{H}}^{2}m_{t}(\pslash'_{t}+\pslash_{t})\gamma^{\mu}C_{0}P_{L}
-m_{u_{H}}^{2}m_{t}\gamma^{\mu}\gamma^{\alpha}C_{\alpha}P_{R}\nonumber\\
&&-m_{u_{H}}^{2}m_{t}\gamma^{\mu}\gamma^{\alpha}C_{\alpha}P_{L}+m_{u_{H}}^{4}\gamma^{\mu}C_{0}P_{L}+m_{u_{H}}^{2}m_{t}^{2}\gamma^{\mu}C_{0}(p'_{t},p_{t},m_{u_{H}},m_{\eta},m_{u_{H}})P_{R}\nonumber\\
&&+\gamma ^{\mu}[m_{t}^{2}B_{1}P_{L}+m_{u_{H}}^{2}B_{1}P_{R}+\frac{1}{2}(m_{t}^{2}+m_{u_{H}}^{2})B_{0}(-p_{t},m_{u_{H}},m_{\eta})\nonumber\\
&&+\frac{1}{2}(m_{t}^{2}+m_{u_{H}}^{2})(m_{t}^{2}+m_{u_{H}}^{2}-m_{\eta}^{2})\frac{\partial}{\partial
p_{t}^{2}}B_{0}-2m_{t}^{2}m_{u_{H}}^{2}\frac{\partial}{\partial
p_{t}^{2}}B_{0}]\}
\end{eqnarray}
\begin{eqnarray}
&&\hat{\Gamma}^{\mu}_{u_{H}\omega^{0}}=\frac{g^{2}g_{s}T^{a}_{\alpha\beta}(V_{Hu})_{i3}^{*}(V_{Hu})_{i3}}{4M_{Z_{H}}^{2}}\frac{i}{16\pi^{2}}\nonumber\\
&&\{m_{u_{H}}^{2}\gamma^{\alpha}\gamma^{\mu}\gamma^{\beta}C_{\alpha\beta}P_{L}+m_{u_{H}}^{2}(\pslash'_{t}+\pslash_{t})\gamma^{\mu}\gamma^{\alpha}C_{\alpha}P_{L}
+m_{t}^{2}\gamma^{\alpha}\gamma^{\mu}\gamma^{\beta}C_{\alpha\beta}P_{R}\nonumber\\
&&+m_{t}^{2}(\pslash'_{t}+\pslash_{t})\gamma^{\mu}\gamma^{\alpha}C_{\alpha}P_{R}
-m_{u_{H}}^{2}m_{t}\gamma^{\alpha}\gamma^{\mu}C_{\alpha}P_{R}-m_{u_{H}}^{2}m_{t}(\pslash'_{t}+\pslash_{t})\gamma^{\mu}C_{0}P_{R}\nonumber\\
&&-m_{u_{H}}^{2}m_{t}\gamma^{\alpha}\gamma^{\mu}C_{\alpha}P_{L}-m_{u_{H}}^{2}m_{t}(\pslash'_{t}+\pslash_{t})\gamma^{\mu}C_{0}P_{L}
-m_{u_{H}}^{2}m_{t}\gamma^{\mu}\gamma^{\alpha}C_{\alpha}P_{R}\nonumber\\
&&-m_{u_{H}}^{2}m_{t}\gamma^{\mu}\gamma^{\alpha}C_{\alpha}P_{L}+m_{u_{H}}^{4}\gamma^{\mu}C_{0}P_{L}+m_{u_{H}}^{2}m_{t}^{2}\gamma^{\mu}C_{0}(p'_{t},p_{t},m_{u_{H}},m_{\omega^{0}},m_{u_{H}})P_{R}\nonumber\\
&&+\gamma^{\mu}[m_{t}^{2}B_{1}P_{L}+m_{u_{H}}^{2}B_{1}P_{R}+\frac{1}{2}(m_{t}^{2}+m_{u_{H}}^{2})B_{0}(-p_{t},m_{u_{H}},m_{\omega^{0}})\nonumber\\
&&+\frac{1}{2}(m_{t}^{2}+m_{u_{H}}^{2})(m_{t}^{2}+m_{u_{H}}^{2}-m_{\omega^{0}}^{2})\frac{\partial}{\partial
p_{t}^{2}}B_{0}-2m_{t}^{2}m_{u_{H}}^{2}\frac{\partial}{\partial
p_{t}^{2}}B_{0}]\}
\end{eqnarray}
\begin{eqnarray}
&&\hat{\Gamma}^{\mu}_{d_{H}\omega^{\pm}}=\frac{g^{2}g_{s}T^{a}_{\alpha\beta}(V_{Hu})_{i3}^{*}(V_{Hu})_{i3}}{2M_{W_{H}}^{2}}\frac{i}{16\pi^{2}}\nonumber\\
&&\{m_{d_{H}}^{2}\gamma^{\alpha}\gamma^{\mu}\gamma^{\beta}C_{\alpha\beta}P_{L}+m_{d_{H}}^{2}(\pslash'_{t}+\pslash_{t})\gamma^{\mu}\gamma^{\alpha}C_{\alpha}P_{L}
+m_{t}^{2}\gamma^{\alpha}\gamma^{\mu}\gamma^{\beta}C_{\alpha\beta}P_{R}\nonumber\\
&&+m_{t}^{2}(\pslash'_{t}+\pslash_{t})\gamma^{\mu}\gamma^{\alpha}C_{\alpha}P_{R}
-m_{d_{H}}^{2}m_{t}\gamma^{\alpha}\gamma^{\mu}C_{\alpha}P_{R}-m_{d_{H}}^{2}m_{t}(\pslash'_{t}+\pslash_{t})\gamma^{\mu}C_{0}P_{R}\nonumber\\
&&-m_{d_{H}}^{2}m_{t}\gamma^{\alpha}\gamma^{\mu}C_{\alpha}P_{L}-m_{d_{H}}^{2}m_{t}(\pslash'_{t}+\pslash_{t})\gamma^{\mu}C_{0}P_{L}
-m_{d_{H}}^{2}m_{t}\gamma^{\mu}\gamma^{\alpha}C_{\alpha}P_{R}\nonumber\\
&&-m_{d_{H}}^{2}m_{t}\gamma^{\mu}\gamma^{\alpha}C_{\alpha}P_{L}+m_{d_{H}}^{4}\gamma^{\mu}C_{0}P_{L}+m_{d_{H}}^{2}m_{t}^{2}\gamma^{\mu}C_{0}P_{R}(p'_{t},p_{t},m_{u_{H}},m_{\omega^{\pm}},m_{u_{H}})\nonumber\\
&&+\gamma^{\mu}[m_{t}^{2}B_{1}P_{L}+m_{d_{H}}^{2}B_{1}P_{R}+\frac{1}{2}(m_{t}^{2}+m_{d_{H}}^{2})B_{0}(-p_{t},m_{u_{H}},m_{\omega^{\pm}})\nonumber\\
&&+\frac{1}{2}(m_{t}^{2}+m_{d_{H}}^{2})(m_{t}^{2}+m_{d_{H}}^{2}-m_{\omega^{\pm}}^{2})\frac{\partial}{\partial
p_{t}^{2}}B_{0}-2m_{t}^{2}m_{d_{H}}^{2}\frac{\partial}{\partial
p_{t}^{2}}B_{0}]\}
\end{eqnarray}
\begin{eqnarray}
&&\hat{\Gamma}^{\mu}_{T^{-}\eta}=\frac{4g'^{2}m^{2}_{t}}{25M^{2}_{A_{H}}}\frac{f^{2}}{v^{2}}[1-\frac{v^{2}}{f^{2}}(\frac{x^{2}_{L}}{2}+\frac{1}{6})]^{2}g_{s}T^{a}_{\alpha\beta}\frac{i}{16\pi^{2}}\nonumber\\
&&\{\gamma^{\alpha}\gamma^{\mu}\gamma^{\beta}C_{\alpha\beta}P_{R}+(\pslash_{t}'+\pslash_{t})\gamma^{\mu}\gamma^{\alpha}C_{\alpha}P_{R}+m^{2}_{T^{-}}\gamma^{\mu}C_{0}(p'_{t},p_{t},m_{T^{-}},m_{\eta},m_{T^{-}})P_{R}\nonumber\\
&&+\gamma_{\mu}[B_{1}P_{L}+\frac{1}{2}B_{0}+\frac{1}{2}(m_{t}^{2}+m^{2}_{T^{-}}-m_{\eta}^{2})\frac{\partial}{\partial
p_{t}^{2}}B_{0}(-p_{t},m_{T^{-}},m_{\eta})]\}
\end{eqnarray}
\begin{eqnarray}
&&\hat{\Gamma}^{\mu}_{T^{-}\omega^{0}}=\frac{g^{2}m_{t}^{2}}{4M^{2}_{Z_{H}}}(\frac{v}{f})^{2}g_{s}T^{a}_{\alpha\beta}\frac{i}{16\pi^{2}}\nonumber\\
&&\{\gamma^{\alpha}\gamma^{\mu}\gamma^{\beta}C_{\alpha\beta}P_{R}+(\pslash_{t}'+\pslash_{t})\gamma^{\mu}\gamma^{\alpha}C_{\alpha}P_{R}+m^{2}_{T^{-}}\gamma^{\mu}C_{0}(p'_{t},p_{t},m_{T^{-}},m_{\omega^{0}},m_{T^{-}})P_{R}\nonumber\\
&&+\gamma_{\mu}[B_{1}P_{L}+\frac{1}{2}B_{0}+\frac{1}{2}(m_{t}^{2}+m^{2}_{T^{-}}-m_{\omega^{0}}^{2})\frac{\partial}{\partial
p_{t}^{2}}B_{0}(-p_{t},m_{T^{-}},m_{\omega^{0}})]\}
\end{eqnarray}
\begin{eqnarray}
&&\hat{\Gamma}^{\mu}_{T^{+}\pi^{0}}=\frac{g^{2}x^{2}_{L}}{4M_{Z}^{2}\cos^{2}\theta}(\frac{v}{f})^{2}g_{s}T^{a}_{\alpha\beta}\frac{i}{16\pi^{2}}\nonumber\\
&&\{m_{T^{+}}^{2}\gamma^{\alpha}\gamma^{\mu}\gamma^{\beta}C_{\alpha\beta}P_{L}+m_{T^{+}}^{2}(\pslash'_{t}+\pslash_{t})\gamma^{\mu}\gamma^{\alpha}C_{\alpha}P_{L}
+m_{t}^{2}\gamma^{\alpha}\gamma^{\mu}\gamma^{\beta}C_{\alpha\beta}P_{R}\nonumber\\
&&+m_{t}^{2}(\pslash'_{t}+\pslash_{t})\gamma^{\mu}\gamma^{\alpha}C_{\alpha}P_{R}
-m_{T^{+}}^{2}m_{t}\gamma^{\alpha}\gamma^{\mu}C_{\alpha}P_{R}-m_{T^{+}}^{2}m_{t}(\pslash'_{t}+\pslash_{t})\gamma^{\mu}C_{0}P_{R}\nonumber\\
&&-m_{T^{+}}^{2}m_{t}\gamma^{\alpha}\gamma^{\mu}C_{\alpha}P_{L}-m_{T^{+}}^{2}m_{t}(\pslash'_{t}+\pslash_{t})\gamma^{\mu}C_{0}P_{L}
-m_{T^{+}}^{2}m_{t}\gamma^{\mu}\gamma^{\alpha}C_{\alpha}P_{R}\nonumber\\
&&-m_{T^{+}}^{2}m_{t}\gamma^{\mu}\gamma^{\alpha}C_{\alpha}P_{L}+m_{T^{+}}^{4}\gamma^{\mu}C_{0}P_{L}+m_{T^{+}}^{2}m_{t}^{2}\gamma^{\mu}C_{0}(p'_{t},p_{t},m_{T^{+}},m_{\pi^{0}},m_{T^{+}})P_{R}\nonumber\\
&&+\gamma^{\mu}[m_{t}^{2}B_{1}P_{L}+m_{T^{+}}^{2}B_{1}P_{R}+\frac{1}{2}(m_{t}^{2}+m_{T^{+}}^{2})B_{0}(-p_{t},m_{T^{+}},m_{\pi^{0}})\nonumber\\
&&+\frac{1}{2}(m_{t}^{2}+m_{T^{+}}^{2})(m_{t}^{2}+m_{T^{+}}^{2}-m_{\pi^{0}}^{2})\frac{\partial}{\partial
p_{t}^{2}}B_{0}+2m_{t}^{2}m_{T^{+}}^{2}\frac{\partial}{\partial
p_{t}^{2}}B_{0}]\nonumber\\
&&+\gamma^{\mu}[m_{t}^{2}B_{1}P_{L}+m_{T^{+}}^{2}B_{1}P_{R}+\frac{1}{2}(m_{t}^{2}+m_{T^{+}}^{2})B_{0}\nonumber\\
&&+\frac{1}{2}(m_{t}^{2}+m_{T^{+}}^{2})(m_{t}^{2}+m_{T^{+}}^{2}-m_{\pi^{0}}^{2})\frac{\partial}{\partial
p_{t}^{2}}B_{0}+2m_{t}^{2}m_{T^{+}}^{2}\frac{\partial}{\partial
p_{t}^{2}}B_{0}]\}
\end{eqnarray}
\begin{eqnarray}
&&\hat{\Gamma}^{\mu}_{T^{+}h}=m_{t}^{2}g_{s}T^{a}_{\alpha\beta}\frac{i}{16\pi^{2}}\nonumber\\
&&\{\frac{c_{\lambda}^{2}}{s_{\lambda}^{2}v^{2}}\gamma^{\alpha}\gamma^{\mu}\gamma^{\beta}C_{\alpha\beta}P_{L}+\frac{c_{\lambda}^{2}}{s_{\lambda}^{2}v^{2}}(\pslash_{t}'+\pslash_{t})\gamma^{\mu}\gamma^{\alpha}C_{\alpha}P_{L}
+\frac{s_{\lambda}^{4}}{f^{2}}\gamma^{\alpha}\gamma^{\mu}\gamma^{\beta}C_{\alpha\beta}P_{R}\nonumber\\
&&+\frac{s_{\lambda}^{4}}{f^{2}}(\pslash_{t}'+\pslash_{t})\gamma^{\mu}\gamma^{\alpha}C_{\alpha}P_{R}
-\frac{c_{\lambda}m_{T^{+}}s_{\lambda}}{vf}\gamma^{\alpha}\gamma^{\mu}C_{\alpha}P_{R}-\frac{c_{\lambda}m_{T^{+}}s_{\lambda}}{vf}(\pslash_{t}'+\pslash_{t})\gamma^{\mu}C_{0}P_{R}\nonumber\\
&&-\frac{c_{\lambda}m_{T^{+}}s_{\lambda}}{vf}\gamma^{\alpha}\gamma^{\mu}C_{\alpha}P_{L}-\frac{c_{\lambda}m_{T^{+}}s_{\lambda}}{vf}(\pslash_{t}'+\pslash_{t})\gamma^{\mu}C_{0}P_{L}
-\frac{c_{\lambda}m_{T^{+}}s_{\lambda}}{vf}\gamma^{\mu}\gamma^{\alpha}C_{\alpha}P_{R}\nonumber\\
&&-\frac{c_{\lambda}m_{T^{+}}s_{\lambda}}{vf}\gamma^{\mu}\gamma^{\alpha}C_{\alpha}P_{L}+\frac{c_{\lambda}^{2}m_{T^{+}}^{2}}{s_{\lambda}^{2}v^{2}}\gamma^{\mu}C_{0}P_{L}+m_{T^{+}}^{2}\frac{s_{\lambda}^{4}}{f^{2}}\gamma^{\mu}C_{0}(p'_{t},p_{t},m_{T^{+}},m_{h},m_{T^{+}})P_{R}\nonumber\\
&&+\gamma^{\mu}[\frac{s_{\lambda}^{4}}{f^{2}}B_{1}P_{L}+\frac{c_{\lambda}^{2}}{s_{\lambda}^{2}v^{2}}B_{1}P_{R}+\frac{1}{2}(\frac{s_{\lambda}^{4}}{f^{2}}+\frac{c_{\lambda}^{2}}{s_{\lambda}^{2}v^{2}})B_{0}(-p_{t},m_{T^{+}},m_{h})\nonumber\\
&&+\frac{1}{2}(\frac{s_{\lambda}^{4}}{f^{2}}+\frac{c_{\lambda}^{2}}{s_{\lambda}^{2}v^{2}})(m_{t}^{2}+m_{T^{+}}^{2}-m_{h}^{2})\frac{\partial}{\partial
p_{t}^{2}}B_{0}-2\frac{s_{\lambda}^{4}}{f^{2}}m_{T^{+}}^{2}\frac{\partial}{\partial
p_{t}^{2}}B_{0}]\}
\end{eqnarray}
\begin{eqnarray}
&&\hat{\Gamma}^{\mu}_{u_{H}A_{H}}=\frac{g'^{2}g_{s}T^{a}_{\alpha\beta}(V_{Hu})_{i3}^{*}(V_{Hu})_{i3}}{100}\frac{i}{16\pi^{2}}
\{2\gamma^{\sigma}\gamma^{\mu}\gamma^{\lambda}C_{\lambda\sigma}P_{L}\nonumber\\
&&+2\gamma^{\mu}P_{L}-\gamma^{\rho}(\pslash'_{t}+\pslash_{t})\gamma^{\mu}\gamma^{\lambda}\gamma_{\rho}C_{\lambda}P_{L}+2m_{u_{H}}^{2}\gamma^{\mu}C_{0}(p'_{t},p_{t},m_{u_{H}},M_{A_{H}},m_{u_{H}})P_{L}\nonumber\\
&&+\gamma^{\mu}[2B_{1}P_{R}+B_{0}+(m_{t}^{2}+m_{u_{H}}^{2}-M_{A_{H}}^{2})\frac{\partial}{\partial
p_{t}^{2}}B_{0}(-p_{t},m_{u_{H}},M_{A_{H}})-P_{L}]\}
\end{eqnarray}
\begin{eqnarray}
&&\hat{\Gamma}^{\mu}_{u_{H}Z_{H}}=\frac{g^{2}g_{s}T^{a}_{\alpha\beta}(V_{Hu})_{i3}^{*}(V_{Hu})_{i3}}{4}\frac{i}{16\pi^{2}}
\{2\gamma^{\sigma}\gamma^{\mu}\gamma^{\lambda}C_{\lambda\sigma}P_{L}\nonumber\\
&&+2\gamma^{\mu}P_{L}-\gamma^{\rho}(\pslash'_{t}+\pslash_{t})\gamma^{\mu}\gamma^{\lambda}\gamma_{\rho}C_{\lambda}P_{L}+2m_{u_{H}}^{2}\gamma^{\mu}C_{0}(p'_{t},p_{t},m_{u_{H}},M_{Z_{H}},m_{u_{H}})P_{L}\nonumber\\
&&+\gamma^{\mu}[2B_{1}P_{R}+B_{0}+(m_{t}^{2}+m_{u_{H}}^{2}-M_{Z_{H}}^{2})\frac{\partial}{\partial
p_{t}^{2}}B_{0}(-p_{t},m_{u_{H}},M_{Z_{H}})-P_{L}]\}
\end{eqnarray}
\begin{eqnarray}
&&\hat{\Gamma}^{\mu}_{d_{H}W_{H}}=\frac{g^{2}g_{s}T^{a}_{\alpha\beta}(V_{Hu})_{i3}^{*}(V_{Hu})_{i3}}{2}\frac{i}{16\pi^{2}}
\{2\gamma^{\sigma}\gamma^{\mu}\gamma^{\lambda}C_{\lambda\sigma}P_{L}\nonumber\\
&&+2\gamma^{\mu}P_{L}-\gamma^{\rho}(\pslash'_{t}+\pslash_{t})\gamma^{\mu}\gamma^{\lambda}\gamma_{\rho}C_{\lambda}P_{L}+2m_{d_{H}}^{2}\gamma^{\mu}C_{0}(p'_{t},p_{t},m_{d_{H}},M_{W_{H}},m_{d_{H}})P_{L}\nonumber\\
&&+\gamma^{\mu}[2B_{1}P_{R}+B_{0}+(m_{t}^{2}+m_{d_{H}}^{2}-M_{W_{H}}^{2})\frac{\partial}{\partial
p_{t}^{2}}B_{0}(-p_{t},m_{d_{H}},M_{W_{H}})-P_{L}]\}
\end{eqnarray}
\begin{eqnarray}
&&\hat{\Gamma}^{\mu}_{T^{-}A_{H}}=\frac{4g'^{2}g_{s}T^{a}_{\alpha\beta}}{25}\frac{i}{16\pi^{2}}\nonumber\\
&&\{2\gamma^{\sigma}\gamma^{\mu}\gamma^{\lambda}x_{L}C_{\lambda\sigma}P_{R}+2x_{L}\gamma^{\mu}P_{R}
+2\gamma^{\lambda}\gamma^{\mu}(\pslash'_{t}+\pslash_{t})x_{L}C_{\lambda}(p'_{t},p_{t},m_{T^{-}},M_{A_{H}},m_{T^{-}})P_{R}\nonumber\\
&&-8m_{T^{-}}x_{L}\sqrt{x_{L}}\frac{v}{f}C^{\mu}-4(p'_{t}+p_{t})^{\mu}m_{T^{-}}x_{L}\sqrt{x_{L}}\frac{v}{f}C_{0}+2m_{T^{-}}^{2}x_{L}\gamma^{\mu}C_{0}P_{R}\nonumber\\
&&+\gamma^{\mu}[2x_{L}B_{1}P_{L}+x_{L}B_{0}+x_{L}(m_{t}^{2}+m_{T^{-}}^{2}-M_{A_{H}}^{2})\frac{\partial}{\partial
p_{t}^{2}}B_{0}-P_{L}\nonumber\\
&&-8x_{L}\sqrt{x_{L}}\frac{v}{f}m_{t}m_{T^{-}}\frac{\partial}{\partial
p_{t}^{2}}B_{0}-\frac{\partial}{\partial
p_{t}^{2}}B_{0}(-p_{t},m_{T^{-}},M_{A_{H}})-x_{L}P_{R}]\}
\end{eqnarray}
\begin{eqnarray}
&&\hat{\Gamma}^{\mu}_{T^{+}Z}=\frac{g^{2}g_{s}T^{a}_{\alpha\beta}x_{L}^{2}}{4\cos^{2}\theta}\frac{v^{2}}{f^{2}}\frac{i}{16\pi^{2}}
\{2\gamma^{\sigma}\gamma^{\mu}\gamma^{\lambda}C_{\lambda\sigma}P_{L}+2\gamma^{\mu}P_{L}\nonumber\\
&&-\gamma^{\rho}(\pslash'_{t}+\pslash_{t})\gamma^{\mu}\gamma^{\lambda}\gamma_{\rho}C_{\lambda}P_{L}
+2m_{T^{+}}^{2}\gamma^{\mu}C_{0}(p'_{t},p_{t},m_{T^{+}},M_{Z},m_{T^{+}})P_{L}\nonumber\\
&&+\gamma^{\mu}[4B_{1}P_{R}+2B_{0}+2(m_{t}^{2}+m_{T^{+}}^{2}-m_{Z}^{2})\frac{\partial}{\partial
p_{t}^{2}}B_{0}(-p_{t},m_{T^{+}},M_{Z})-2P_{L}]\}
\end{eqnarray}

(II)Renormalization fermion propagator
\begin{figure}[htbp]
\scalebox{0.4}{\epsfig{file=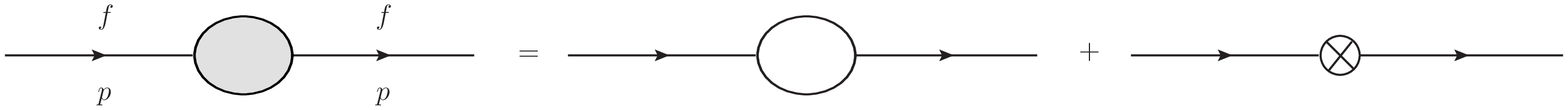}}
\end{figure}
\begin{eqnarray*}
-i\hat{\Sigma}^{f}(p)=-i\Sigma^{f}(p)+(-i\delta\Sigma^{f}(p))
\end{eqnarray*}
where
\begin{eqnarray*}
&&\Sigma^{f}(p)=m_{f}\Sigma^{f}_{S}(p^{2})+\pslash P_{L}\Sigma^{f}_{L}(p^{2})+\pslash P_{R}\Sigma^{f}_{R}(p^{2})\\
&&\delta\Sigma^{f}(p)=\delta m_{f}+m_{f}\frac{1}{2}\delta Z_{L}^{f}+m_{f}\frac{1}{2}\delta Z_{R}^{f}-\pslash P_{L}\delta Z_{L}^{f}-\pslash P_{R}\delta Z_{R}^{f}\\
&&\delta
m_{f}=-m_{f}Re[\Sigma^{f}_{S}(m_{f}^{2})+\frac{1}{2}\Sigma^{f}_{L}(m_{f}^{2})+\frac{1}{2}\Sigma^{f}_{R}(m_{f}^{2})]\\
&&\delta
Z_{L}^{f}=Re\Sigma^{f}_{L}(m_{f}^{2})+m_{f}^{2}\frac{\partial}{\partial
p^{2}}Re[\Sigma^{f}_{L}(p^{2})+\Sigma^{f}_{R}(p^{2})+2\Sigma^{f}_{S}(p^{2})]|_{p^{2}=m_{f}^{2}}\\
&&\delta
Z_{R}^{f}=Re\Sigma^{f}_{R}(m_{f}^{2})+m_{f}^{2}\frac{\partial}{\partial
p^{2}}Re[\Sigma^{f}_{L}(p^{2})+\Sigma^{f}_{R}(p^{2})+2\Sigma^{f}_{S}(p^{2})]|_{p^{2}=m_{f}^{2}}
\end{eqnarray*}
\begin{eqnarray*}
\hat{\Sigma}^{t}=\hat{\Sigma}^{t}(\eta)+\hat{\Sigma}^{t}(\omega^{0})+\hat{\Sigma}^{t}(\omega^{\pm})+\hat{\Sigma}^{t}(\pi^{0})
+\hat{\Sigma}^{t}(h)+\hat{\Sigma}^{t}(A_{H})+\hat{\Sigma}^{t}(Z_{H})+\hat{\Sigma}^{t}(W_{H}^{\pm})+\hat{\Sigma}^{t}(Z)
\end{eqnarray*}
\begin{eqnarray}
&&-i\hat{\Sigma}^{\eta
u_{H}}(p)=-\frac{g'^{2}(V_{Hu})_{i3}^{*}(V_{Hu})_{i3}}{100M^{2}_{A_{H}}}\frac{i}{16\pi^{2}}\nonumber\\
&&\{m^{2}_{u_{H}}\pslash_{t}B_{1}P_{L}+m^{2}_{t}\pslash_{t}B_{1}P_{R}+m_{t}m^{2}_{u_{H}}B_{0}(-p_{t},m_{u_{H}},m_{\eta})\nonumber\\
&&-m_{t}m_{u_{H}}^{2}B_{0}+m_{t}^{3}\frac{\partial}{\partial
p_{t}^{2}}(m_{u_{H}}^{2}B_{1}+m_{t}^{2}B_{1}+2m_{u_{H}}^{2}B_{0})\nonumber\\
&&-\pslash_{t}P_{L}[m_{u_{H}}^{2}B_{1}+m_{t}^{2}\frac{\partial}{\partial
p_{t}^{2}}(m_{u_{H}}^{2}B_{1}+m_{t}^{2}B_{1}+2m_{u_{H}}^{2}B_{0})]\nonumber\\
&&-\pslash_{t}P_{R}[m_{u_{H}}^{2}B_{1}+m_{t}^{2}\frac{\partial}{\partial
p_{t}^{2}}(m_{u_{H}}^{2}B_{1}+m_{t}^{2}B_{1}+2m_{u_{H}}^{2}B_{0})]\}
\end{eqnarray}
\begin{eqnarray}
&&-i\hat{\Sigma}^{\omega^{0}
u_{H}}(p)=-\frac{g'^{2}(V_{Hu})_{i3}^{*}(V_{Hu})_{i3}}{100M^{2}_{A_{H}}}\frac{i}{16\pi^{2}}\nonumber\\
&&\{m^{2}_{u_{H}}\pslash_{t}B_{1}P_{L}+m^{2}_{t}\pslash_{t}B_{1}P_{R}+m_{t}m^{2}_{u_{H}}B_{0}(-p_{t},m_{u_{H}},m_{\omega^{0}})\nonumber\\
&&-m_{t}m_{u_{H}}^{2}B_{0}+m_{t}^{3}\frac{\partial}{\partial
p_{t}^{2}}(m_{u_{H}}^{2}B_{1}+m_{t}^{2}B_{1}+2m_{u_{H}}^{2}B_{0})\nonumber\\
&&-\pslash_{t}P_{L}[m_{u_{H}}^{2}B_{1}+m_{t}^{2}\frac{\partial}{\partial
p_{t}^{2}}(m_{u_{H}}^{2}B_{1}+m_{t}^{2}B_{1}+2m_{u_{H}}^{2}B_{0})]\nonumber\\
&&-\pslash_{t}P_{R}[m_{u_{H}}^{2}B_{1}+m_{t}^{2}\frac{\partial}{\partial
p_{t}^{2}}(m_{u_{H}}^{2}B_{1}+m_{t}^{2}B_{1}+2m_{u_{H}}^{2}B_{0})]\}
\end{eqnarray}
\begin{eqnarray}
&&-i\hat{\Sigma}^{\omega^{\pm}
d_{H}}(p)=-\frac{g'^{2}(V_{Hu})_{i3}^{*}(V_{Hu})_{i3}}{100M^{2}_{A_{H}}}\frac{i}{16\pi^{2}}\nonumber\\
&&\{m^{2}_{d_{H}}\pslash_{t}B_{1}P_{L}+m^{2}_{t}\pslash_{t}B_{1}P_{R}+m_{t}m^{2}_{d_{H}}B_{0}(-p_{t},m_{d_{H}},m_{\omega^{\pm}})\nonumber\\
&&-m_{t}m_{d_{H}}^{2}B_{0}+m_{t}^{3}\frac{\partial}{\partial
p_{t}^{2}}(m_{d_{H}}^{2}B_{1}+m_{t}^{2}B_{1}+2m_{d_{H}}^{2}B_{0})\nonumber\\
&&-\pslash_{t}P_{L}[m_{d_{H}}^{2}B_{1}+m_{t}^{2}\frac{\partial}{\partial
p_{t}^{2}}(m_{d_{H}}^{2}B_{1}+m_{t}^{2}B_{1}+2m_{d_{H}}^{2}B_{0})]\nonumber\\
&&-\pslash_{t}P_{R}[m_{d_{H}}^{2}B_{1}+m_{t}^{2}\frac{\partial}{\partial
p_{t}^{2}}(m_{d_{H}}^{2}B_{1}+m_{t}^{2}B_{1}+2m_{d_{H}}^{2}B_{0})]\}
\end{eqnarray}
\begin{eqnarray}
&&-i\hat{\Sigma}^{\eta
T^{-}}(p)=-(\frac{2g'm_{t}f}{5M_{A_{H}}v})^{2}\frac{i}{16\pi^{2}}[1-\frac{v^{2}}{f^{2}}(\frac{x^{2}_{L}}{2}+\frac{1}{6})]^{2}
[\pslash_{t}B_{1}P_{R}\nonumber\\
&&+m_{t}^{3}\frac{\partial}{\partial
p_{t}^{2}}B_{1}-\pslash_{t}P_{L}m_{t}^{2}\frac{\partial}{\partial
p_{t}^{2}}B_{1}-\pslash_{t}P_{R}(B_{1}+m_{t}^{2}\frac{\partial}{\partial
p_{t}^{2}}B_{1})(-p_{t},m_{T^{-}},m_{\eta})]
\end{eqnarray}
\begin{eqnarray}
&&-i\hat{\Sigma}^{\omega_{0}
T^{-}}(p)=-(\frac{gm_{t}f}{2M_{Z_{H}}v})^{2}\frac{i}{16\pi^{2}}\nonumber\\
&&[\pslash_{t}B_{1}P_{R} +m_{t}^{3}\frac{\partial}{\partial
p_{t}^{2}}B_{1}-\pslash_{t}P_{L}m_{t}^{2}\frac{\partial}{\partial
p_{t}^{2}}B_{1}-\pslash_{t}P_{R}B_{1}(-p_{t},m_{T^{-}},m_{\omega_{0}})]
\end{eqnarray}
\begin{eqnarray}
&&-i\hat{\Sigma}^{\pi^{0}
T^{+}}(p)=-\frac{g^{2}x_{L}^{2}v^{2}}{4M^{2}_{Z}\cos^{2}\theta
f^{2}}\frac{i}{16\pi^{2}}\nonumber\\
&&\{m^{2}_{T^{+}}\pslash_{t}B_{1}P_{L}+m^{2}_{t}\pslash_{t}B_{1}P_{R}+m_{t}m^{2}_{T^{+}}B_{0}(-p_{t},m_{T^{+}},m_{\pi^{0}})\nonumber\\
&&-m_{t}m_{T^{+}}^{2}B_{0}+m_{t}^{3}\frac{\partial}{\partial
p_{t}^{2}}(m_{T^{+}}^{2}B_{1}+m_{t}^{2}B_{1}+2m_{T^{+}}^{2}B_{0})\nonumber\\
&&-\pslash_{t}P_{L}[m_{T^{+}}^{2}B_{1}+m_{t}^{2}\frac{\partial}{\partial
p_{t}^{2}}(m_{T^{+}}^{2}B_{1}+m_{t}^{2}B_{1}+2m_{T^{+}}^{2}B_{0})]\nonumber\\
&&-\pslash_{t}P_{R}[m_{T^{+}}^{2}B_{1}+m_{t}^{2}\frac{\partial}{\partial
p_{t}^{2}}(m_{T^{+}}^{2}B_{1}+m_{t}^{2}B_{1}+2m_{T^{+}}^{2}B_{0})]\}
\end{eqnarray}
\begin{eqnarray}
&&-i\hat{\Sigma}^{h T^{+}}(p)=m_{t}^{2}\frac{i}{16\pi^{2}}
\{\frac{x_{L}}{(1-x_{L})v^{2}}\pslash_{t}B_{1}P_{L}-\frac{(1-x_{L})^{2}}{f^{2}}\pslash_{t}B_{1}P_{R}\nonumber\\
&&-\frac{s_{\lambda}\sqrt{x_{L}}m_{T^{+}}}{fv}B_{0}+\frac{s_{\lambda}\sqrt{x_{L}}m_{T^{+}}}{fv}B_{0}(-p_{t},m_{T^{+}},m_{h})\nonumber\\
&&+m_{t}\frac{\partial}{\partial
p_{t}^{2}}[\frac{x_{L}m_{t}^{2}}{(1-x_{L})v^{2}}B_{1}-\frac{(1-x_{L})^{2}m_{t}^{2}}{f^{2}}B_{1}-\frac{2s_{\lambda}\sqrt{x_{L}}m_{T^{+}}m_{t}}{fv}B_{0}]\\
&&-\pslash_{t}P_{L}[\frac{x_{L}}{(1-x_{L})v^{2}}B_{1}+\frac{\partial}{\partial
p_{t}^{2}}[\frac{x_{L}m_{t}^{2}}{(1-x_{L})v^{2}}B_{1}-\frac{(1-x_{L})^{2}m_{t}^{2}}{f^{2}}B_{1}-\frac{2s_{\lambda}\sqrt{x_{L}}m_{T^{+}}m_{t}}{fv}B_{0}]\nonumber\\
&&-\pslash_{t}P_{R}[\frac{x_{L}}{(1-x_{L})v^{2}}B_{1}+\frac{\partial}{\partial
p_{t}^{2}}[\frac{x_{L}m_{t}^{2}}{(1-x_{L})v^{2}}B_{1}-\frac{(1-x_{L})^{2}m_{t}^{2}}{f^{2}}B_{1}-\frac{2s_{\lambda}\sqrt{x_{L}}m_{T^{+}}m_{t}}{fv}B_{0}]\}\nonumber
\end{eqnarray}
\begin{eqnarray}
&&-i\hat{\Sigma}^{A_{H}
u_{H}}(p)=-\frac{g'^{2}(V_{Hu})_{i3}^{*}(V_{Hu})_{i3}}{100}\frac{i}{16\pi^{2}}
[2\pslash_{t}(B_{1}+\frac{1}{2})P_{L}\nonumber\\
&&+2m_{t}^{3}\frac{\partial}{\partial
p_{t}^{2}}B_{1}-2\pslash_{t}P_{L}(B_{1}+\frac{1}{2}+m_{t}^{2}\frac{\partial}{\partial
p_{t}^{2}}B_{1})-2\pslash_{t}P_{R}m_{t}^{2}\frac{\partial}{\partial
p_{t}^{2}}B_{1}(-p_{t},m_{u_{H}},M_{A_{H}})]
\end{eqnarray}
\begin{eqnarray}
&&-i\hat{\Sigma}^{Z_{H}
u_{H}}(p)=-\frac{g^{2}(V_{Hu})_{i3}^{*}(V_{Hu})_{i3}}{4}\frac{i}{16\pi^{2}}
[2\pslash_{t}(B_{1}+\frac{1}{2})P_{L}\nonumber\\
&&+2m_{t}^{3}\frac{\partial}{\partial
p_{t}^{2}}B_{1}-2\pslash_{t}P_{L}(B_{1}+\frac{1}{2}+m_{t}^{2}\frac{\partial}{\partial
p_{t}^{2}}B_{1})-2\pslash_{t}P_{R}m_{t}^{2}\frac{\partial}{\partial
p_{t}^{2}}B_{1}(-p_{t},m_{u_{H}},M_{Z_{H}})]
\end{eqnarray}
\begin{eqnarray}
&&-i\hat{\Sigma}^{W_{H}
d_{H}}(p)=-\frac{g^{2}(V_{Hu})_{i3}^{*}(V_{Hu})_{i3}}{2}\frac{i}{16\pi^{2}}
[2\pslash_{t}(B_{1}+\frac{1}{2})P_{L}\nonumber\\
&&+2m_{t}^{3}\frac{\partial}{\partial
p_{t}^{2}}B_{1}-2\pslash_{t}P_{L}(B_{1}+\frac{1}{2}+m_{t}^{2}\frac{\partial}{\partial
p_{t}^{2}}B_{1})-2\pslash_{t}P_{R}m_{t}^{2}\frac{\partial}{\partial
p_{t}^{2}}B_{1}(-p_{t},m_{d_{H}},M_{W_{H}})]
\end{eqnarray}
\begin{eqnarray}
&&-i\hat{\Sigma}^{Z
T^{+}}(p)=-\frac{g^{2}x_{L}^{2}v^{2}}{4\cos^{2}\theta
f^{2}}\frac{i}{16\pi^{2}}[2\pslash_{t}(B_{1}+\frac{1}{2})P_{L}\nonumber\\
&&+2m_{t}^{3}\frac{\partial}{\partial
p_{t}^{2}}B_{1}-2\pslash_{t}P_{L}(B_{1}+\frac{1}{2}+m_{t}^{2}\frac{\partial}{\partial
p_{t}^{2}}B_{1})-2\pslash_{t}P_{R}m_{t}^{2}\frac{\partial}{\partial
p_{t}^{2}}B_{1}(-p_{t},m_{T^{+}},M_{Z})]
\end{eqnarray}
\begin{eqnarray}
&&-i\hat{\Sigma}^{A_{H}
T^{-}}(p)=-\frac{4g'^{2}}{25}\frac{i}{16\pi^{2}}
\{2x_{L}^{2}\frac{v^{2}}{f^{2}}\pslash_{t}(B_{1}+\frac{1}{2})P_{L}+2x_{L}\pslash_{t}(B_{1}+\frac{1}{2})P_{R}\nonumber\\
&&+2m_{t}^{3}\frac{\partial}{\partial
p_{t}^{2}}[x_{L}^{2}\frac{v^{2}}{f^{2}}B_{1}+x_{L}B_{1}+4x_{L}\sqrt{x_{L}}\frac{v}{f}\frac{m_{T^{-}}}{m_{t}}B_{0}(-p_{t},m_{T^{-}},M_{A_{H}})]\nonumber\\
&&-2\pslash_{t}P_{L}[x_{L}^{2}\frac{v^{2}}{f^{2}}(B_{1}+\frac{1}{2})+m_{t}^{2}\frac{\partial}{\partial
p_{t}^{2}}(x_{L}^{2}\frac{v^{2}}{f^{2}}B_{1}+x_{L}B_{1}+4x_{L}\sqrt{x_{L}}\frac{v}{f}\frac{m_{T^{-}}}{m_{t}}B_{0})]\nonumber\\
&&-2\pslash_{t}P_{R}[x_{L}^{2}\frac{v^{2}}{f^{2}}(B_{1}+\frac{1}{2})+m_{t}^{2}\frac{\partial}{\partial
p_{t}^{2}}(x_{L}^{2}\frac{v^{2}}{f^{2}}B_{1}+x_{L}B_{1}+4x_{L}\sqrt{x_{L}}\frac{v}{f}\frac{m_{T^{-}}}{m_{t}}B_{0})]\}
\end{eqnarray}

\end{document}